\documentclass[useAMS,usenatbib]{mn2e}
\usepackage{graphicx,natbib}
\usepackage{psfrag}
\usepackage{amssymb}
\usepackage{float}

\def\aV{\mbox{$\rm A_V$}}

\title[Galactic anticentre star clusters]{Towards a census of the Galactic anticentre star clusters - III: tracing the spiral structure in the Outer Disk}

\author[D. Camargo, E. Bica and C. Bonatto]{D. Camargo$^1$, E. Bica$^1$ and C. Bonatto$^1$\\
$^1$ Departamento de Astronomia, Universidade Federal do Rio Grande do Sul, 
Av. Bento Gon\c{c}alves 9500\\
Porto Alegre 91501-970, RS, Brazil}

\begin{document}

\pagerange{\pageref{firstpage}--\pageref{lastpage}}

\maketitle

\label{firstpage}

\begin{abstract}

In this paper we investigate the nature of 27 star cluster candidates, most of them
projected towards the Galactic anticentre. We derive fundamental parameters for 20
confirmed clusters, among these 7 are new identifications. Four of the remaining are
uncertain cases that require deeper photometry to establish their nature, and 4 are
probably field fluctuations. In addition, we provide a partial census of the
open clusters towards the Galactic anticentre. We also include in this study some
interesting objects outside the anticentre region, in the second and third Galactic
quadrants, mainly in the Perseus and Outer arms. These clusters confirm the extension
of the Outer arm along the third quadrant. We also point out that the embedded cluster
FSR 486, at a distance of $7.2\pm1.3$ kpc from de Sun, is projected on the line of sight
of the Local Group irregular dwarf galaxy IC 10. Thus, part of the unusual properties
of IC 10 may be explained by a Galactic contamination. We point out the importance of
embedded clusters in tracing the spiral structure.

\end{abstract}

\begin{keywords}
({\it Galaxy}:) open clusters and associations:general; {\it Galaxy}: disc; {\it Galaxy}: structure; {\it Galaxy}: catalogues; {\it galaxies}: dwarf galaxies;
\end{keywords}

\section{Introduction}
\label{Intro}
\citet{Trumpler30b} was the pioneer in using star clusters as tracers of the Galactic structure. Since then, se\-veral works have used open clusters (OCs) as probes to trace the structure and understand the dynamics of our Galaxy, especially the disk \citep{Janes82, Friel95, Kroupa02a, Bonatto06a, Piskunov06, Bobylev07, Vazquez08}.

Given the importance of OCs to increase our know\-ledge of the Galaxy, numerous efforts have been made to expand the OC sample and improve the accuracy of the derived parameters. On the observational point of view, se\-veral catalogues and surveys were compiled \citep{Alter70, Lynga87, Dias02, Dutra03, Bica03a, Bica03b, Bica05, Kharchenko05a,Kharchenko05b, Froebrich07, Koposov08, Glushkova10}. In this sense, we have contributed significantly to increase the number of clusters with parameters derived towards the Galactic anticentre \citep[][- thereafter Papers I, II, and III]{Camargo10, Camargo11, Camargo12}.

To improve the accuracy of the cluster parameters, our group has developed a series of tools and procedures \citep{Bonatto07b, Bica08, Bonatto12}. In particular, a field decontamination algorithm (Sect.~\ref{sec:3}) that is essential to better determine the cluster members and differentiate physical systems from field fluctuations, mainly in the analysis of poorly populated objects, those in crowded fields or embedded in nebulae \citep{Bonatto09, Bonatto11, Camargo09, Camargo11, Camargo12, Pavani11, Gunes12}.

In the present work we investigate the nature of some subsamples of star cluster candidates. In the first one, we test the nature of cluster candidates towards the Galactic anticentre, in the sector $160^\circ\,\leq\,\ell\,\leq 200^\circ$. Ten of them are NGC cluster candidates. These objects have recently been included in a catalogue of overlooked NGC OCs \citep{Tadross11}. We employ a field decontamination procedure to uncover the intrinsic cluster CMDs. Another subsample is candidate clusters from \citet{Froebrich07}. We also analyse some interesting objects outside the anticentre region. In addition, we provide new clusters disco\-vered by ourselves (CBB 10 to CBB 16) following the CBB series (Papers II and III). The search for new clusters are made by eye on WISE, 2MASS, and XDSS images. Then, we condensed our recent contributions to the anticentre clusters into a catalogue (Papers I, II, and III). Using this catalogue we investigate properties of both anticentre clusters and the 2nd and 3rd quadrants.

\begin{figure}
\resizebox{\hsize}{!}{\includegraphics{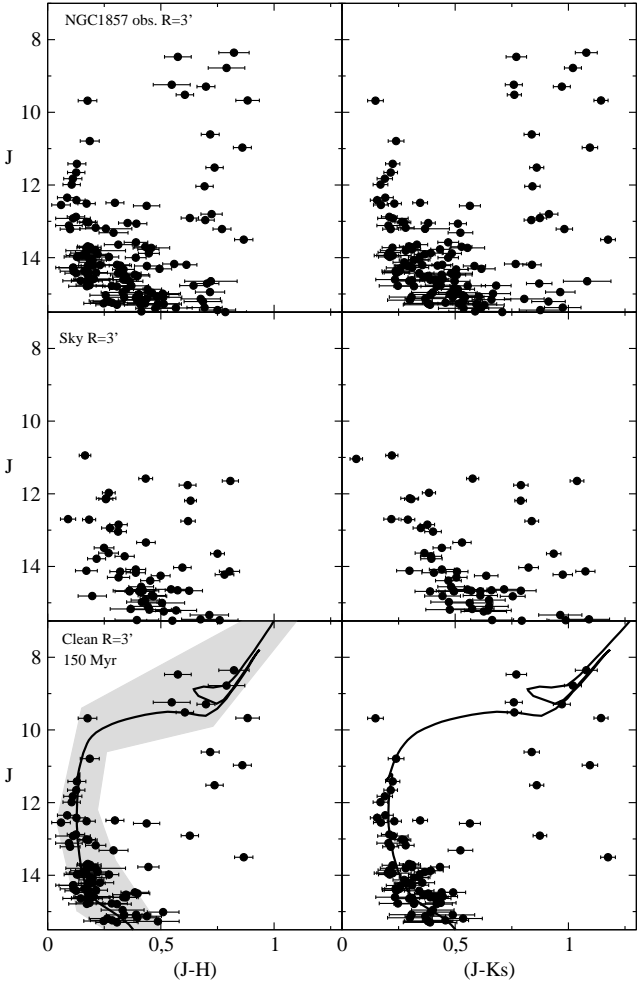}}
\caption[]{2MASS CMDs extracted from the $R=3'$ region of NGC 1857. \textit{Top panels}: observed CMDs $J\times(J-H)$ (\textit{left}) and $J\times(J-K_s)$ (\textit{right}). \textit{Middle}: equal area comparison field. \textit{Bottom}: field-star decontaminated CMDs fitted with 150 Myr Padova isochrones (solid line). The colour-magnitude filter used to isolate cluster stars is shown as a shaded region.}
\label{fig:1}
\end{figure}

\begin{figure}
\resizebox{\hsize}{!}{\includegraphics{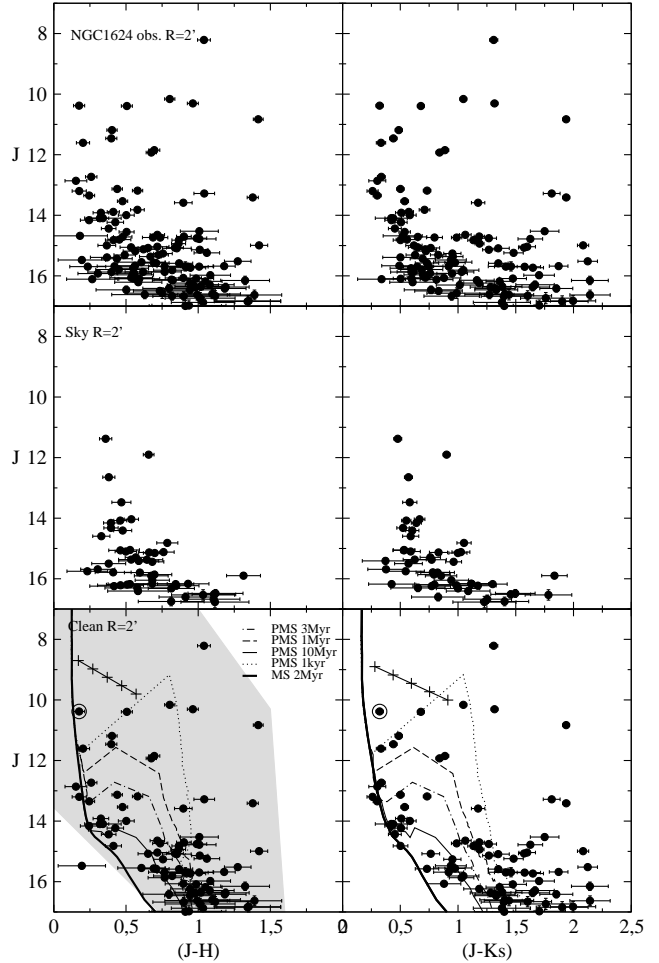}}
\caption[]{Same as Fig.~\ref{fig:1} for NGC 1624. The circle on the decontaminated CMDs indicates an O star. The decontaminated CMD of NGC 1624 was fitted with Padova isochrones (2 Myr) for the MS and Siess (0.1, 1, 3, and 10 Myr) for PMS stars. We also show the reddening vector for $A_v=0$ to 5.}
\label{fig:02}
\end{figure}

The paper is organized as follows. In Sect. \ref{sec:3} we present the 2MASS photometry and describe the methods and tools employed in the CMD analyses, especially the field star decontamination algorithm. Sect. \ref{sec:4} is dedicated to the discussion of the methods and tools used for the analysis of the cluster structure. In Sect. \ref{sec:5} we present the results of the analyses of cluster candidates, and derive astrophysical parameters (\textit{age, reddening, distance, core and cluster radii}) of the confirmed OCs. In Sect. \ref{sec:7} we discuss the results. Finally, in Sect. \ref{sec:8} we provide the concluding remarks.

\begin{table}
\centering
{\footnotesize
\caption{General data of the present star clusters or candidates.}
\label{tab1}
\renewcommand{\tabcolsep}{3.0mm}
\renewcommand{\arraystretch}{1,2}
\begin{tabular}{lrrrrr}
\hline
\hline
Target&$\alpha(2000)$&$\delta(2000)$&$\ell$&$b$\\
&(h\,m\,s)&$(^{\circ}\,^{\prime}\,^{\prime\prime})$&$(^{\circ})$&$(^{\circ})$ \\
($1$)&($2$)&($3$)&($4$)&($5$)\\
\hline

CBB 10 &4:05:58&51:06:07&151.05&-0.83\\
CBB 11 &4:40:27&50:28:30&155.33&2.59\\
CBB 12 &4:41:14&50:28:00&155.42&2.68\\
CBB 13 &6:27:39&12:32:39&199.08&0.51\\
CBB 14 &6:54:34&00:14:58&213.06&0.81\\
CBB 15 &6:54:46&00:23:38&212.96&0.92\\

CBB 16 &7:18:34&-13:12:42&227.75&-0.108\\

FSR486 &0:20:21&59:19:05&118.96&-3.31\\

FSR831 &6:28:50&33:38:46&180.63&10.95\\

FSR835 &4:53:59&20:18:28&180.90&-13.93\\

FSR843 &7:00:37&35:04:38&181.67&16.93\\

FSR851 &5:14:39&19:48:01&183.85&-10.88\\

FSR909 &6:15:46&19:00:49&192.03&1.04\\

FSR913 &5:25:26&11:35:20&192.67&-12.57\\

FSR1099 &6:34:27&-3:49:37&214.41&-5.52\\
FSR1234 &7:31:10&-15:25:46&231.16&1.53\\

NGC1624 &4:40:38&50:27:45&155.36&2.61\\
NGC1807 &5:10:46&16:30:30&186.09&-13.495\\
NGC1857 &5:20:04&39:18:38&168.43&1.23\\
NGC2026 &5:43:12&20:08:00&187.23&-05.059\\
NGC2039 &5:44:00&8:41:30&197.58&-10.274\\
NGC2063 &5:46:43&8:46:54&197.29&-10.766\\
NGC2165 &6:11:0.4&51:40:36&162.16&15.13\\
NGC2224 &6:27:28&12:36:55&198.99&0.51\\
NGC2248 &6:34:35&26:18:16&187.54&8.24\\
NGC2331 &7:06:59&27:15:42&189.728&15.218\\
NGC2666 &8:49:47&44:42:12&175.92&39.278\\
\hline
\end{tabular}
\begin{list}{Table Notes.}
\item Cols. $2-3$: Central coordinates. Cols. $4-5$: Corresponding Galactic coordinates.
\end{list}
}
\end{table}

\begin{figure}
\resizebox{\hsize}{!}{\includegraphics{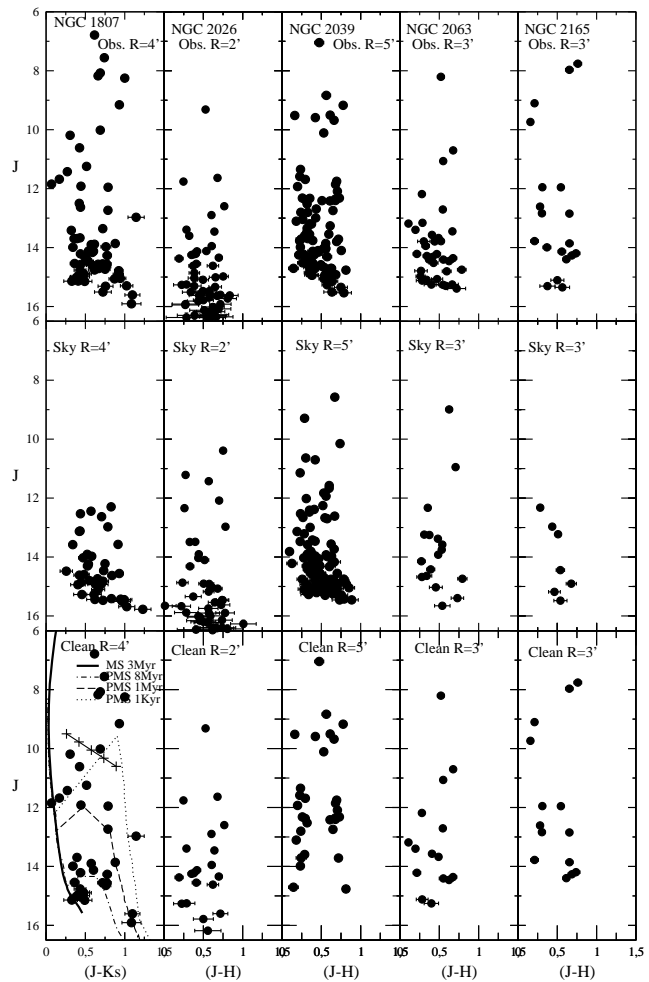}}
\caption[]{2MASS CMDs of the candidates NGC 1807, NGC 2026, NGC 2039, NGC 2063, and NGC 2165. \textit{Top panels}: observed CMDs $J\times(J-H)$. \textit{Middle}: equal area comparison field. \textit{Bottom}: field-star decontaminated CMDs. The decontaminated CMD of NGC 1807 was fitted with Padova isochrones (3 Myr) for the MS and Siess (0.1, 1, and 8 Myr) for PMS stars. We also show the reddening vector for $A_v=0$ to 5.}
\label{fig:03}
\end{figure}

\begin{figure}
\resizebox{\hsize}{!}{\includegraphics{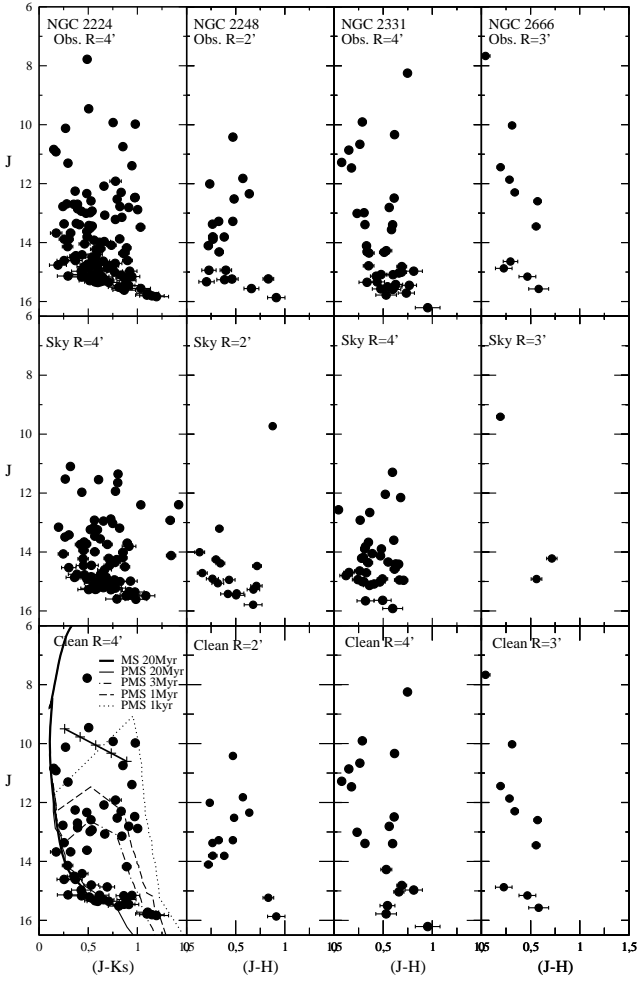}}
\caption[]{Same as Fig.~\ref{fig:03} for NGC 2248, NGC 2331, and NGC 2666, for NGC 2224 is shown the $J\times(J-K_s)$ CMDs. }
\label{fig:04}
\end{figure}

\section{2MASS photometry and analytical tools}
\label{sec:3}

We analyse the cluster candidates with 2MASS\footnote{The Two Micron All Sky Survey, available at \textit{www..ipac.caltech.edu/2mass/releases/allsky/}} photo\-metry \citep{Skrutskie06} in the $J$, $H$ and $K_{s}$ bands, extracted in circular concentric regions centred on the coordinates given in Table~\ref{tab4} using VizieR\footnote{http://vizier.u-strasbg.fr/viz-bin/VizieR?-source=II/246.}. Large extraction areas are essential for a consistent field star decontamination (Sect.~\ref{sec:3.1}) and to obtain RDPs (Sect.~\ref{sec:4}) with a high contrast relative to the background. In addition, 2MASS provides an all-sky coverage with the spatial and photome\-tric uniformity required for high star-count statistics. As a photometric quality constraint, the 2MASS extractions were restricted to stars with errors in $J$, $H$ and $K_{s}$ smaller than $0.2$ mag.

The extraction radius was chosen by visual ins\-petion on the WISE image and taking into account the RDP, in the sense that the profile must become relatively stable in the outer region. The wide extraction area is needed to provide the required statistics, in terms of magnitude and colours, for a consistent field star decontamination. 

\subsection{Field-star decontamination}
\label{sec:3.1}

There is evident contamination by field stars in the observed CMDs of the present cluster candidates. Therefore, we applied the field-star decontamination procedure given by \citet{Bonatto07a} to uncover the intrinsic CMD morpho\-logy from the foreground/background stars. 

The decontamination algorithm is described in detail in \citet{Bonatto07b} and \citet{Bica08}, therefore we provide a brief description. The tool divides the CMD into a 3D grid of cells with axes along the $J$ magnitude and the ($J-H$) and ($J-K_s$) colours, computing the expected number-density of field stars in a given cell based on the number of comparison field stars with magnitude and colours compatible with those of the cell. Subsequently, it subtracts the expected number of field stars from each cell. Typical cell dimensions are $\Delta{J}=1.0$, and  ${\Delta(J-H)}={\Delta(J-K_{s})}=0.2$, which are large enough to allow sufficient star-count statistics in individual cells and small enough to maintain the morphology of the CMD evolutionary sequences.  

The comparison field used in the decontamination procedure is chosen by considering the field in the cluster neighbourhood. In general, large field a\-reas are required to ensure statistical representativeness of field stars. The decontamination constraints more the parameters derived, especially for low-latitude OCs and ECs.

We argue that it is not an easy task to establish the nature of faint objects and/or those in crowded field, thus the field star decontamination algorithm plays an important role in disentangling physical CMD sequences from field fluctuations.

\begin{figure}
\resizebox{\hsize}{!}{\includegraphics{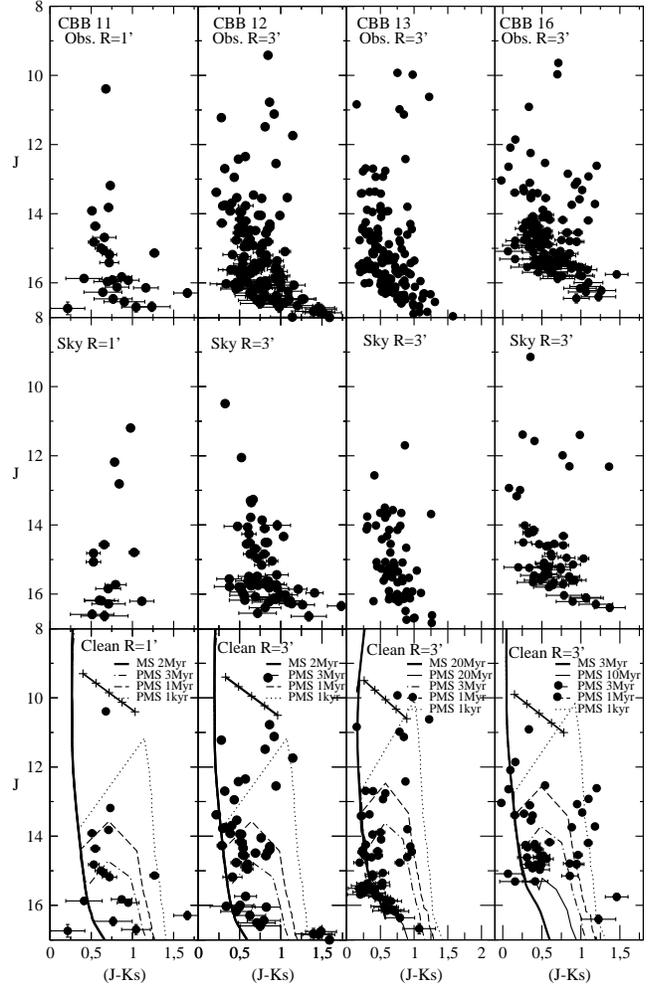}}
\caption[]{ 2MASS CMDs extracted from the central region of CBB 11, CBB 12, CBB 13, and CBB 16. \textit{Top panels}: observed CMDs $J\times(J-K_s)$. \textit{Middle panels}: equal area comparison field. \textit{Bottom panels}: field-star decontaminated CMDs fitted with Padova isochrones for the MS and Siess isochrones for PMS stars. We also give the reddening vector for $A_v=0$ to 5.}
\label{fig:05}
\end{figure}

\subsection{Fundamental parameters}
\label{sec:3.2}

\begin{figure}
\resizebox{\hsize}{!}{\includegraphics{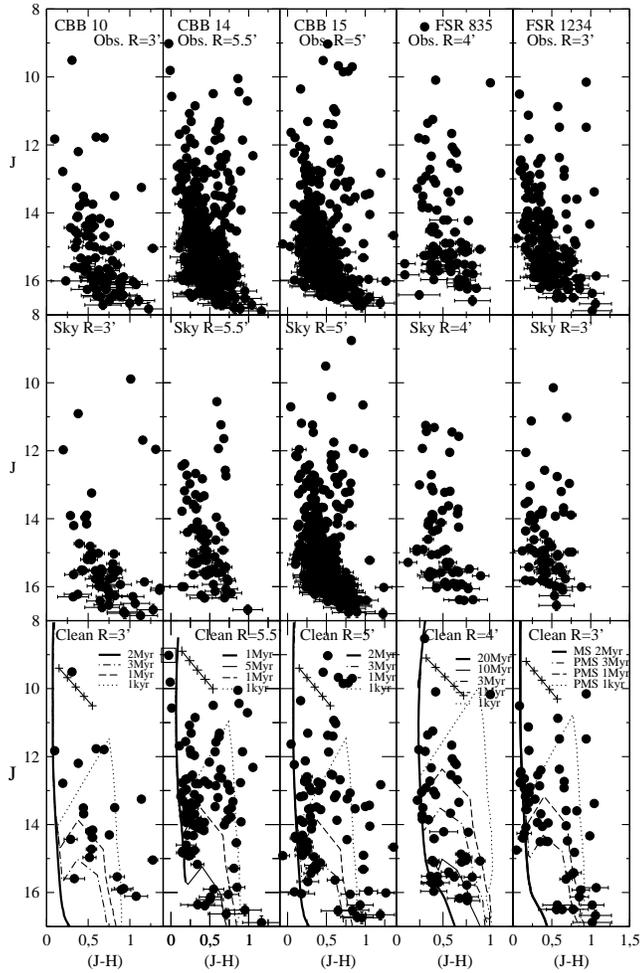}}
\caption[]{Same as Fig.~\ref{fig:03} for CBB 15, CBB 10, CBB 14, FSR 835, and FSR 1234. The square in CBB 14's subtracted diagram indicates a Be star.}
\label{fig:06}
\end{figure}

\begin{figure}
\resizebox{\hsize}{!}{\includegraphics{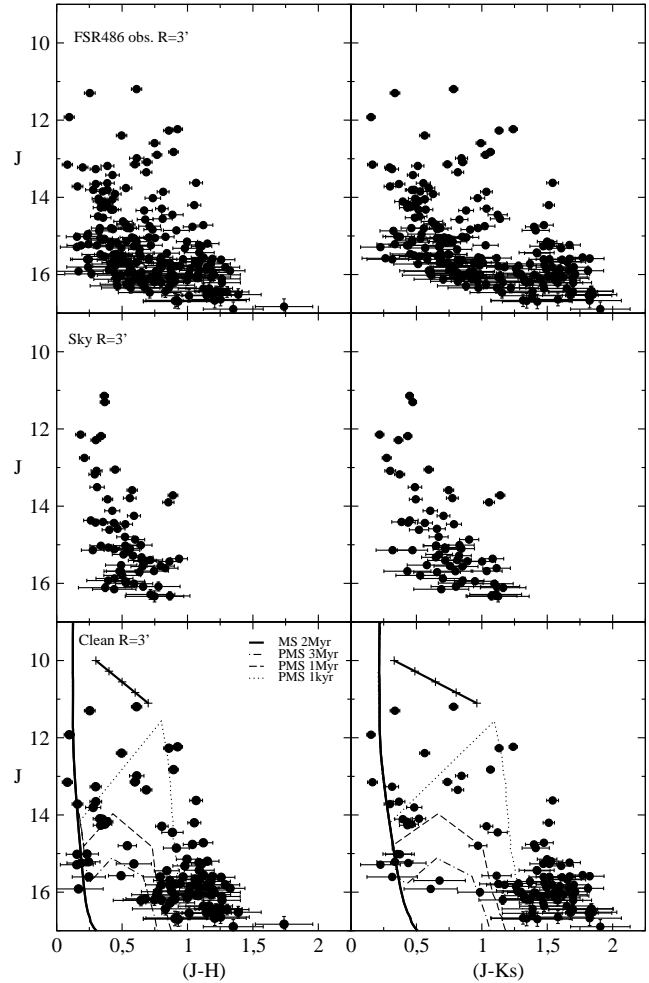}}
\caption[]{Same as Fig.~\ref{fig:1} for FSR 486.}
\label{fig:07}
\end{figure}

\begin{figure}
\resizebox{\hsize}{!}{\includegraphics{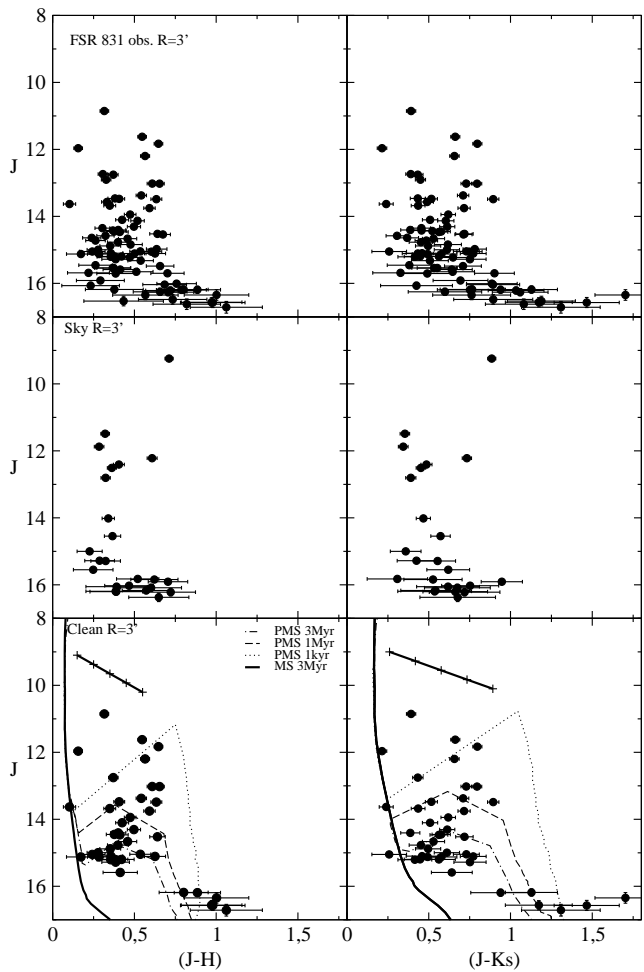}}
\caption[]{Same as Fig.~\ref{fig:1} for FSR 831.}
\label{fig:08}
\end{figure}

\begin{figure}
\resizebox{\hsize}{!}{\includegraphics{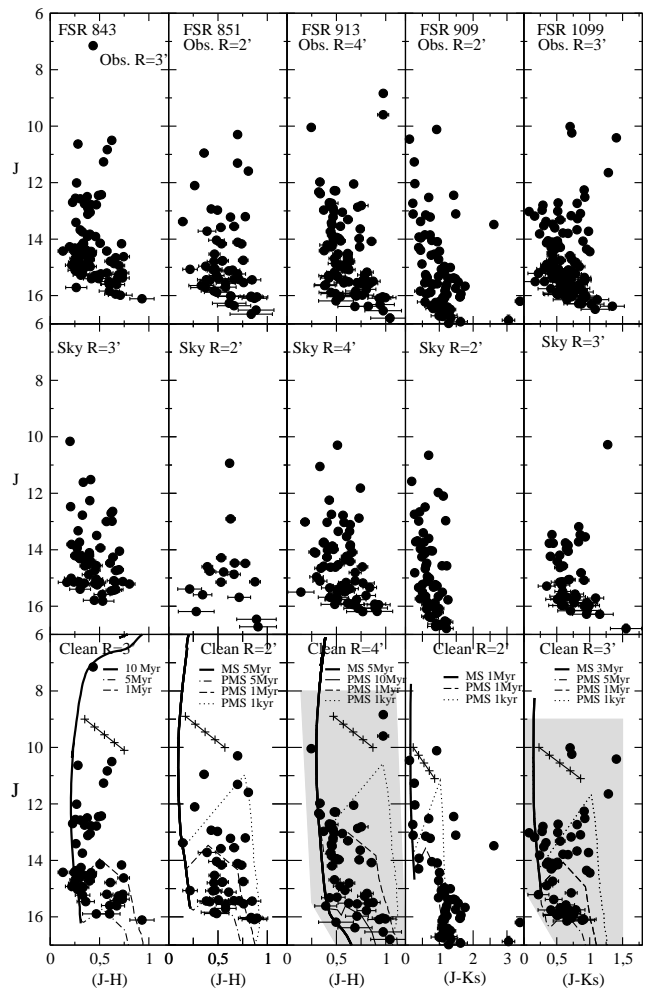}}
\caption[]{Same as Fig.~\ref{fig:03} for FSR 843, FSR 851, FSR 913, FSR 909, and FSR 1099.}
\label{fig:09}
\end{figure}

We estimate the fundamental parameters using the decontaminated CMD morphology fitted by eye with Padova isochrones \citep{Marigo08}. The derived parameters are the observed distance modulus $(m-M)_{J}$ and reddening $E(J-H)$, which converts to $E(B-V)$ and $A_{V}$ with the relations $A_{J}/{A_{V}}=0.276$, $A_{H}/{A_{V}}=0.176$, $A_{K_{s}}/{A_{V}}=0.118$, $A_{J}=2.76\times{E(J-H)}$  and $E(J-H)=0.33\times{E(B-V)}$ \citep{Dutra02}. We assume a constant total-to-selective absorption ratio $R_{V}=3.1$ \citep{Cardelli89}. The analysis is based on both CMDs, $J\times(J-H)$ and $J\times(J-K_s)$, but for brevity we show only one of them for most objects. For clusters presenting stars with IR-excess we provide $J\times(J-K_s)$ CMDs.  Decontaminated CMDs are shown in Figs.~\ref{fig:1} to \ref{fig:09}. The parameter errors have been estimated by displacing the best-fitting isochrone in colour and magnitude to the limiting point where the fit remains acceptable.

\begin{table*}
{\footnotesize
\begin{center}
\caption{Derived fundamental parameters for confirmed star clusters in the present study.}
\renewcommand{\tabcolsep}{2.9mm}
\renewcommand{\arraystretch}{1.3}
\begin{tabular}{lrrrrrrrr}
\hline
\hline
Cluster&phase&$A_V$&Age&$d_{\odot}$&$R_{GC}$&$x_{GC}$&$y_{GC}$&$z_{GC}$\\
&&(mag)&(Myr)&(kpc)&(kpc)&(kpc)&(kpc)&(kpc)\\
($1$)&($2$)&($3$)&($4$)&($5$)&($6$)&($7$)&($8$)&($9$) \\
\hline
\multicolumn{8}{c}{Confirmed OCs}\\
\hline

CBB 10 &EC&$2.18\pm0.2$&$2\pm1$&$7.2\pm1.2$&$13.98\pm1.2$&$-13.54\pm1.2$&$3.49\pm0.33$&$-0.1\pm0.01$\\
CBB 11 &EC&$2.58\pm0.2$&$2\pm1$&$6.0\pm1.0$&$12.9\pm1.0$&$-12.65\pm1.0$&$+2.49\pm0.24$&$0.27\pm0.03$\\
CBB 12 &EC&$2.18\pm0.2$&$2\pm1$&$6.0\pm1.0$&$12.9\pm1.0$&$-12.69\pm1.0$&$+2.46\pm0.25$&$0.34\pm0.03$\\
CBB 13 &EC\textsuperscript{t}&$2.78\pm0.2$&$20\pm10$&$3.5\pm0.5$&$10.6\pm0.5$&$-10.53\pm0.5$&$-1.17\pm0.11$&$0.07\pm0.01$\\

CBB 14 &EC&$1.39\pm0.2$&$1\pm1$&$6.7\pm1.0$&$13.3\pm1.0$&$-12.8\pm1.0$&$-3.63\pm0.4$&$0.1\pm0.02$\\

CBB 15 &EC&$2.18\pm0.2$&$2\pm1$&$7.2\pm1.5$&$13.85\pm1.5$&$-13.28\pm1.5$&$-3.93\pm0.5$&$0.12\pm0.01$\\

CBB 16 &EC&$1.29\pm0.2$&$3\pm2$&$4.25\pm0.75$&$10.56\pm0.7$&$-10.08\pm0.7$&$-3.14\pm0.3$&$-0.01\pm0.01$\\

FSR486 &EC&$2.18\pm0.2$&$2\pm1$&$7.2\pm1.3$&$12.4\pm1.3$&$-10.71\pm1.2$&$6.31\pm0.80$&$-0.42\pm0.04$\\
FSR831 &EC\textsuperscript{t}&$2.28\pm0.2$&$2\pm1$&$6.2\pm1.2$&$13.37\pm1.2$&$-13.33\pm1.2$&$-0.03\pm0.01$&$1.12\pm0.11$\\
FSR835 &EC\textsuperscript{t}&$3.47\pm0.2$&$20\pm10$&$3.07\pm0.4$&$10.22\pm0.4$&$-10.20\pm0.4$&$-0.05\pm0.01$&$-0.74\pm0.07$\\
FSR843 &EC\textsuperscript{t}&$3.47\pm0.2$&$10\pm5$&$6.71\pm0.6$&$13.8\pm0.9$&$-13.64\pm0.9$&$-0.19\pm0.02$&$+1.95\pm0.19$\\

FSR851 &EC\textsuperscript{t}&$1.98\pm0.2$&$5\pm3$&$6.16\pm1.1$&$13.3\pm1.2$&$-13.26\pm1.2$&$-0.41\pm0.04$&$-1.16\pm0.11$\\

FSR909 &EC&$1.98\pm0.2$&$1\pm1$&$6.45\pm1.0$&$13.5\pm1.0$&$-13.53\pm1.0$&$-1.34\pm0.15$&$0.12\pm0.01$\\

FSR913 &EC\textsuperscript{t}&$4.46\pm0.2$&$5\pm3$&$3.56\pm0.5$&$10.7\pm0.6$&$-10.61\pm0.6$&$-0.76\pm0.08$&$-0.78\pm0.07$\\

FSR1099 &EC\textsuperscript{t}&$1.98\pm0.2$&$3\pm2$&$8.12\pm1.5$&$14.6\pm1.5$&$-13.9\pm1.5$&$-4.57\pm0.44$&$-0.78\pm0.08$\\

FSR1234 &EC&$2.48\pm0.2$&$2\pm1$&$5.0\pm0.8$&$11.1\pm0.9$&$-10.38\pm0.8$&$-3.92\pm0.37$&$0.13\pm0.01$\\

NGC1624 &EC&$2.58\pm0.2$&$2\pm1$&$6.0\pm1.0$&$12.9\pm1.0$&$-12.65\pm1.0$&$+2.49\pm0.24$&$0.27\pm0.03$\\
NGC1807 &EC&$2.78\pm0.2$&$3\pm2$&$2.79\pm1.0$&$9.94\pm1.0$&$-9.92\pm1.0$&$-0.29\pm0.03$&$-0.65\pm0.06$\\
NGC1857 &OC&$1.98\pm0.2$&$150\pm50$&$2.8\pm0.5$&$10.0\pm0.5$&$-9.99\pm0.5$&$+0.56\pm0.05$&$+0.06\pm0.01$\\
NGC2224 &EC\textsuperscript{t}&$2.28\pm0.2$&$20\pm20$&$2.36\pm0.5$&$9.48\pm0.5$&$-9.45\pm0.5$&$-0.77\pm0.1$&$0.02\pm0.01$\\

\hline
\end{tabular}
\begin{list}{Table Notes.}
\item Col. 2: evolutionary phase - EC means embedded cluster, EC\textsuperscript{t} embedded cluster in probable phase transition, and OC open cluster; Col. 3: $A_V$ in the cluster's central region. Col. 4: age, from 2MASS photometry. Col. 5: distance from the Sun. Col. 6: $R_{GC}$ calculated using $R_{\odot}=7.2$ kpc for the distance of the Sun to the Galactic centre \citep{Bica06}. Cols. 7 - 9: Galactocentric components.
\end{list}
\label{tab4}
\end{center}
}
\end{table*}

\begin{table*}
{\footnotesize
\begin{center}
\caption{Structural parameters for clusters in the current sample.}
\renewcommand{\tabcolsep}{3.5mm}
\renewcommand{\arraystretch}{1.3}
\begin{tabular}{lrrrrrrrr}
\hline
\hline
Cluster&$(1')$&$\sigma_{0K}$&$R_{core}$&$R_{RDP}$&$\sigma_{0K}$&$R_{core}$&$R_{RDP}$&${\Delta}R$\\
&($pc$)&($*\,pc^{-2}$)&($pc$)&($pc$)&($*\,\arcmin^{-2}$)&($\arcmin$)&($\arcmin$)&($\arcmin$)\\
($1$)&($2$)&($3$)&($4$)&($5$)&($6$)&($7$)&($8$)&($9$)\\
\hline
FSR486 &$2.09$&$2.8\pm1.0$&$1.75\pm0.48$&$17.1\pm3.1$&$12.11\pm4.4$&$0.84\pm0.23$&$8.2\pm3.0$&$20-40$\\
FSR913 &$1.03$&$8.48\pm1.2$&$0.52\pm$0.04&$5.2\pm1.5$&$9.0\pm1.3$&$0.51\pm0.04$&$5.0\pm1.5$&$20-60$\\

FSR1099 &$2.35$&$1.27\pm0.5$&$1.34\pm0.47$&$14.1\pm3.5$&$7.0\pm2.8$&$0.57\pm0.2$&$6.0\pm1.5$&$20-60$\\

FSR1234 &$1.46$&$3.1\pm1.5$&$1.0\pm0.27$&$10.2\pm2.9$&$6.65\pm2.2$&$0.74\pm0.19$&$7.0\pm2.0$&$20-50$\\

NGC1624 &$1.73$&$15.97\pm6.4$&$0.64\pm0.17$&$6.9\pm1.7$&$47.8\pm19.1$&$0.37\pm0.1$&$4.0\pm1.0$&$20-40$\\
NGC1857 &$0.82$&$10.8\pm5$&$0.88\pm0.25$&$16.4\pm4.1$&$7.26\pm3.4$&$1.08\pm0.3$&$20.0\pm5.0$&$40-80$\\

\hline
\end{tabular}
\begin{list}{Table Notes.}
\item Col. 2: arcmin to parsec scale. To minimise degrees of freedom in RDP fits with the King-like profile (see text), $\sigma_{bg}$ was kept fixed (measured in the respective comparison fields) while $\sigma_{0}$ and $R_{core}$ were allowed to vary. Col. 11: comparison field ring.
\end{list}
\label{tab5}
\end{center}
}
\end{table*}

\subsection{Cluster structure}
\label{sec:4}

\begin{figure}
\resizebox{\hsize}{!}{\includegraphics{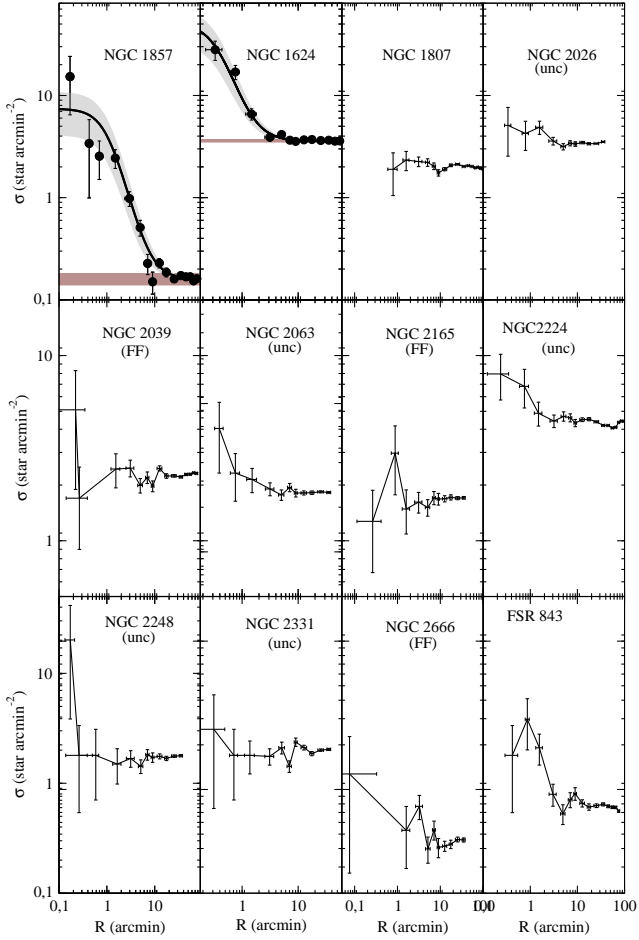}}
\caption[]{Radial density profiles for the confirmed clusters NGC 1857, NGC 1624, NGC 1807, and FSR 843, for the uncertain cases NGC 2026, NGC 2063, NGC 2224, NGC 2248, and NGC 2331, and for the probable field fluctuations. Brown horizontal shaded region: stellar background level measured in the comparison field. Gray regions: $1\sigma$ King fit uncertainty.}
\label{fig:10}
\end{figure}

\begin{figure}
\resizebox{\hsize}{!}{\includegraphics{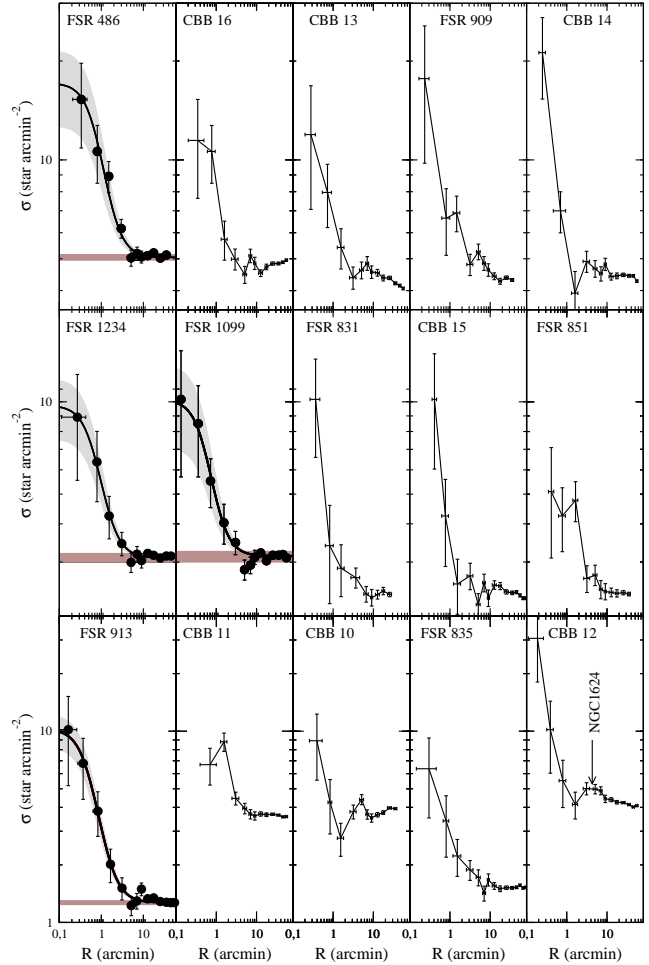}}
\caption[]{Radial density profiles for the remaining confirmed clusters.}
\label{fig:11}
\end{figure}

The structure of each cluster candidate is analysed by means of the stellar radial density profile (RDP), which is built with stars selected after applying the respective colour magnitude (CM) filter to the observed photometry.
The CM filter isolates the probable cluster sequences excluding stars with different colours, enhancing the RDP contrast relative to the background \citep[e.g.][and references therein]{Bonatto07a}. Nevertheless, it is expected that residual field stars with colours similar to those of the cluster remain in the CM filter. The effect of this residual contamination in the intrinsic RDP depends on the relative densities of field and cluster stars. The practical effect of applying the CM filter in the cluster sequences is a significant enhancement of the contrast of the RDP with respect to the background. Ho\-wever, for ECs with stars presenting IR-excess the filter frequently selects all stars in the decontaminated CMDs of the cluster central region, but the filter is applied in the overall observed photometry. To avoid oversampling near the centre and undersampling for large radii, the RDPs were built by counting stars in concentric rings of increasing width with distance to the centre. The number and width of rings are adjusted so that the resul\-ting RDPs present adequate spatial resolution with moderate $1\sigma$ Poisson errors. The R coordinate (and respective uncertainty) of a given ring within the RDP corresponds to the average distance to the cluster centre (and standard deviation) computed for the stars within the ring. The CM filters for clusters with RDPs following a King's law \citep{King62} are shown in Figs.~\ref{fig:1}, \ref{fig:02}, and \ref{fig:09} as the shaded area superimposed on the field-star decontaminated CMD. RDPs are shown in  Figs.~\ref{fig:10} and \ref{fig:11}.

Following previous work \citep[e.g.,][]{Bica11, Bonatto10, Bonatto11a} we fit the RDPs with the function $\sigma(R)=\sigma_{bg}+\sigma_{0}/(1+(R/R_{core})^{2}$, where $\sigma_{bg}$ is the residual background surface density of stars, $\sigma_{0}$ is the central density of stars and $R_{core}$ is the core radius. The cluster radius ($R_{RDP}$) and uncertainty can be estimated by considering the fluctuations of the RDPs with respect to the residual background. $R_{RDP}$ corresponds to the distance from the centre where both RDP and comparison field become statistically indistinguishable. This function, applied to star counts, is similar to that introduced by King to des\-cribe the surface-brightness profiles in the central parts of GCs. To minimise degrees of freedom $\sigma_{bg}$ is measured in the comparison field and kept fixed.

\section{Results}
\label{sec:5}

The present objects are classified into three groups, accor\-ding to the photometric and RDP analyses. The group of confirmed open clusters includes the objects for which it was possible to derive fundamental parameters (Tables \ref{tab8} and \ref{tab9}). The objects classified as uncertain cases present less defined decontaminated CMD sequences. These objects generally present irregular RDPs. However, ECs may also present these irregularities. The objects whose evolutionary sequences were not recognized were classified as field fluctuations.

\subsection{Galactic anticentre candidates}
\label{sec:5.1}

We adopt as Galactic anticentre the sector $160^\circ\,\leq\,\ell\,\leq 200^\circ$ analysed in our previous works (Papers I, II, and III). The objects of the present sample in this sector are FSR 831, FSR 835, FSR 843, FSR 851, FSR 909, FSR 913, 12 NGC cluster candidates, and the newly discovered CBB 13. 

The decontamination procedure and structural analysis show that most of the present NGC OC candidates are probably field fluctuations.  
We were able to derive the fundamental parameters for NGC 1624 \citep[see also][]{Jose11}, NGC 1807 \citep{Balaguer04}, and NGC 1857 (Tab. \ref{tab4}) and structural parameters for NGC 1857 (Tab. \ref{tab5}).  We classify NGC 2026, NGC 2063, NGC 2248, and NGC 2331 as uncertain and suggest deeper photometry to determine their nature. The remaining NGC objects are probably field fluctuations.

Fig.~\ref{fig:1} shows the $J\times(J-H)$ and $J\times(J-K_s)$ CMDs extracted from a region $R=3'$ centred on the coordinates of the confirmed cluster NGC 1857 (top-panels). The middle panels show the comparison field corresponding to a ring with the same area as the central region. In the bottom panels we show the decontaminated CMDs fitted with 150 Myr Padova isochrones \citep{Marigo08}.  Figs.~\ref{fig:03} and \ref{fig:04} show the CMDs for the remaining NGC OC candidates.

We also derive fundamental parameters for CBB 13, FSR 831, FSR 835, FSR 843, FSR 851, FSR 909, and FSR 913 (Tab.~\ref{tab4} and Figs.~\ref{fig:05}, \ref{fig:06}, and \ref{fig:09}) and structural parameters for FSR 913 (Tab.~\ref{tab4} and Fig.~\ref{fig:11}).

\subsection{Additional objects outside the anticentre slice}
\label{sec:5.2}

We also analyse 10 overdensities outside the anticentre region, 6 of them discovered in the present work. These clusters might help uncover the spiral structure in the Outer Galactic disk. CBB 10, CBB 11, CBB 12, NGC 1624, and FSR 486 are distant clusters located in the Outer arm in the second quadrant. CBB 11, CBB 12, and NGC 1624 are related to the H II region Sh2-212 that is probably deve\-loping a sequential star formation triggered by massive stars within NGC 1624 \citep{Deharveng08}. CBB 14, CBB 15, CBB 16, FSR 1234, and FSR 1099 help trace the spiral structure (Perseus and Outer arms) in the third quadrant.

We derive fundamental parameters for the clusters (Table \ref{tab4}) and the structural parameters for FSR 486, FSR 1099, FSR 1234, and NGC 1624 (Table \ref{tab5}).
CMDs for the newly found clusters are shown in Figs.~\ref{fig:05} and \ref{fig:06}, for NGC 1624 in Fig.~\ref{fig:04}, and the remaining clusters in Figs.~\ref{fig:06} and \ref{fig:09}.

\section{Discussion}
\label{sec:7}
\subsection{General}

The present work is part of a series of papers dedicated to the analysis of the Galactic anticentre star clusters in the sector $160^\circ\,\leq\,\ell\,\leq 200^\circ$ (Papers I, II, and III). The result of this effort is a catalogue of OCs and ECs in the Galactic anticentre direction (Tables \ref{tab4} and \ref{tab8} - online material). Besides the present sample we also include in this study some additional interesting objects outside the anticentre region.

The fundamental parameters derived for the catalogued star clusters are shown in Table \ref{tab8} and the structural parameters are shown in Table \ref{tab9}.

Recent observations indicate that nebulae may present several cluster-forming \textit{close clumps} with ongoing star formation, mainly in filamentary structures \citep{Gutermuth08, Camargo11, Feigelson11, Camargo12, Fernandes12}. After the gas expulsion phase, the surviving clumps may form an association of clusters or merge forming a massive single cluster (Paper III). As a result, it is possible that young clusters present CMDs with considerable age spread and large RDPs with several bumps and dips.

On the other hand, \citet{Parmentier12} point out that, for a spherical gas clump, the cluster central region forms first, and then the star formation propagates outwardly. In this sense, several young clusters present a bimodal star formation \citep{Pfalzner09}. This bimodality appears to be dependent on both radius and time. Most clusters probably undergo a substructured phase with RDPs that do not follow a King law and CMDs with some age spread. Some clusters of the present sample, such as CBB 13 and NGC 2224, may present this effect, with a significant spread in age. Other clusters, especially ECs, present RDPs concentrated in the central region, such as CBB 12, CBB 13, CBB 14, and CBB 15.

\subsection{FSR 486 and the Dwarf galaxy IC 10}
\label{sec:7.2}

\begin{figure*}
\begin{center}
   \includegraphics[scale=0.385,angle=0,viewport=0 0 635 635,clip]{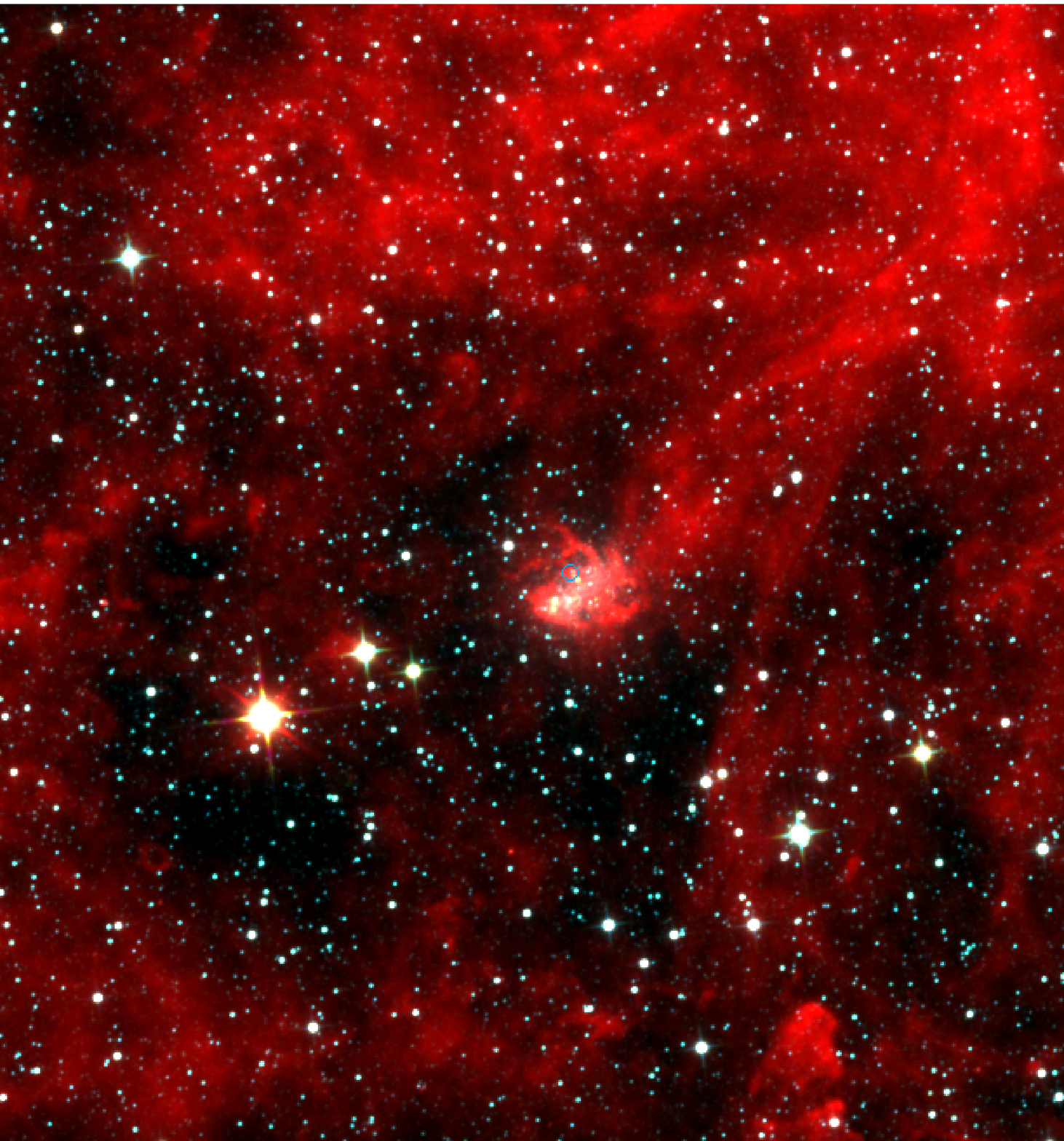}
   \includegraphics[scale=0.28,angle=0,viewport=0 0 900 900,clip]{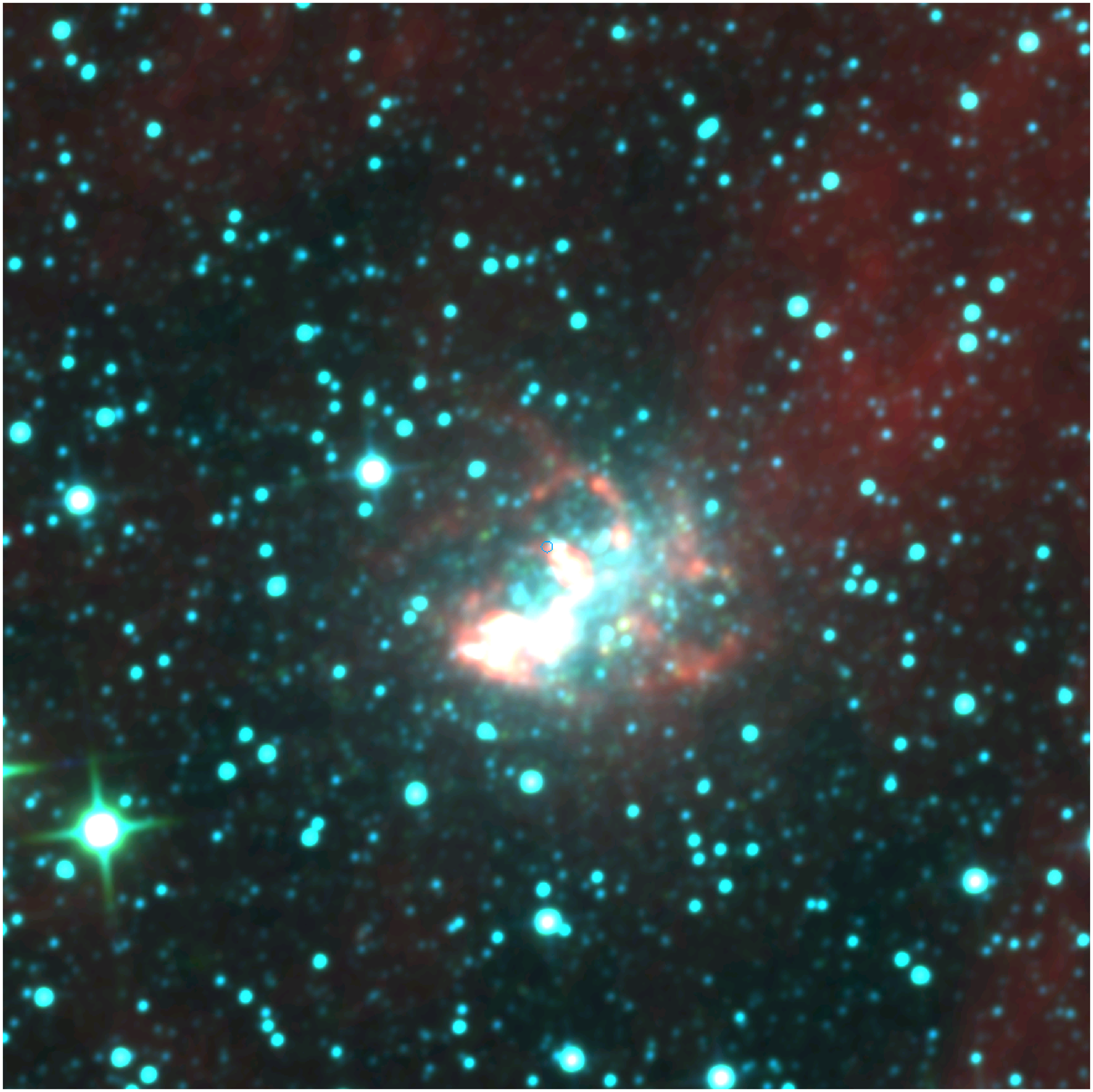}
\caption[]{Left: WISE image composition ($60'\times60'$) of dust emission centred on FSR 486 and IC 10. Right: composite colour image ($20'\times20'$) of the WISE bands centred on FSR 486. All stars in both panels belong to the embedded cluster FSR 486 or to the Galactic field.}
\label{fig:12}
\end{center}
\end{figure*}

\begin{figure}
\begin{center}
\includegraphics[scale=0.38,angle=0]{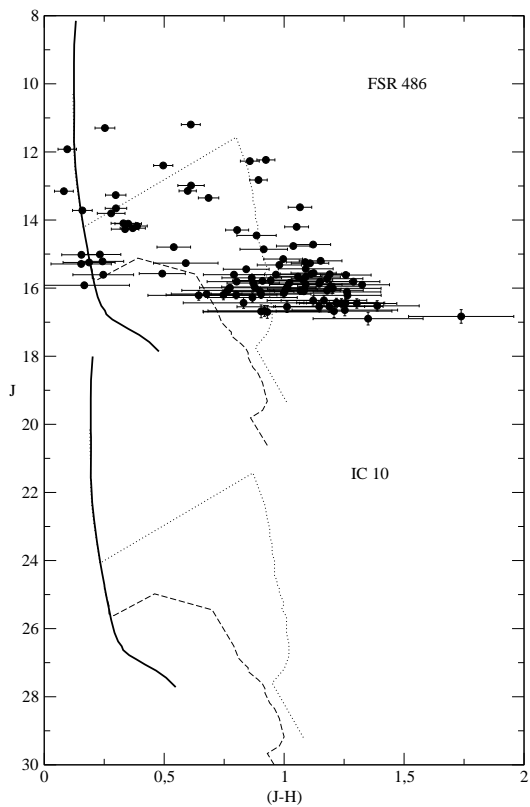}
\caption[]{Decontaminated CMD of FSR 486. Lines are Padova and Siess isochrones representing our fit for FSR 486, and the \citet{Gonçalves12} parameters converted to isochrones for the Dwarf galaxy IC 10.}
\label{fig:13}
\end{center}
\end{figure}

Our CMD and RDP (Figs.~\ref{fig:07} and \ref{fig:11}) indicate the star cluster nature of FSR 486 \citep{Froebrich07}. We point out that FSR 486 is projected towards the Local Group star forming galaxy IC 10 (Fig.~\ref{fig:12}), which has been studied with HST data \citep{Sanna10}. Comparing the CMDs and considering the different photometric bands of both studies, the CMD sequences of the star cluster and galaxy hardly overlap (Fig.~\ref{fig:13}). The set of Wolf-Rayet and O stars from IC 10 are too faint to be present in the 2MASS photometry \footnote{see http://simbad.u-strasbg.fr/simbad/sim-fcoo}. Hence, our CMDs (Fig.~\ref{fig:07}) consist exclusively of FSR 486 stars. The derived distance modulus to IC 10 in previous works is larger than 24 mag \citep{Borissova00, Hunter01, Sanna08, Gonçalves12}. However, we derive a distance modulus of $\approx14$ mag for FSR 486. We also point out the possible presence of an HII region where the EC FSR 486 is forming. IC 10 presents unusual gas kinematics, star formation and physical parameters \citep[][and references therein]{Cohen79, Bolatto00, Sanna10}, but part of this scenario may be a contamination by the Galaxy.

\subsection{Galactic distribution}
\label{sec:6.2}

\begin{figure*}
\begin{center}
   \includegraphics[scale=0.65,angle=0]{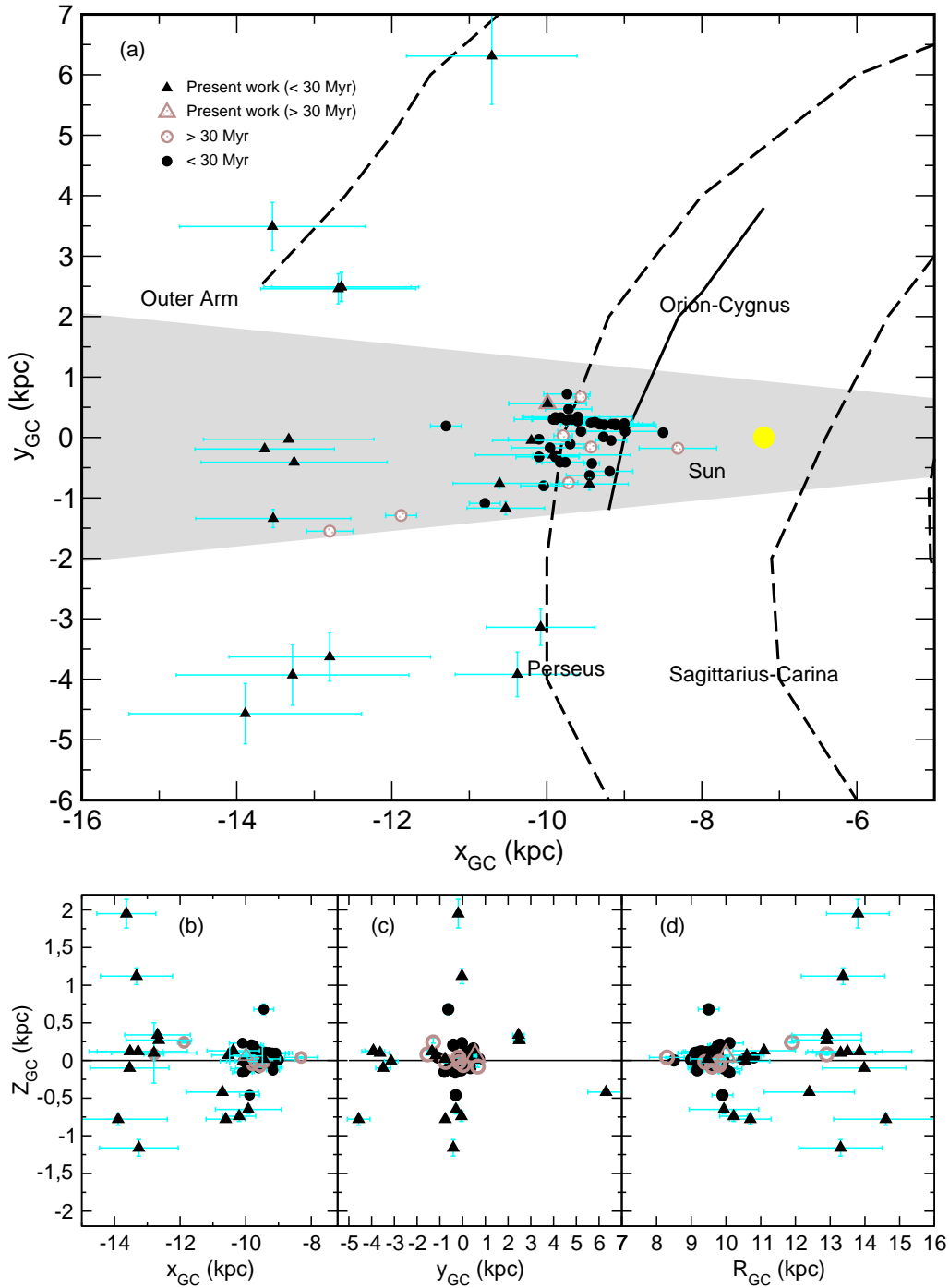}
   \caption[]{Spatial distribution of the star clusters in Papers I, II, III (circle) and the present ones (triangles). The shaded area indicates the sector $160^\circ\,\leq\,\ell\,\leq 200^\circ$.}
\label{fig:14}
\end{center}
\end{figure*}

Fig.~\ref{fig:14} shows the spatial distribution of the clusters in Papers I, II, and III and in the present work on a schematic view of the Galactic disk \citep{Momany06}.
We classify the present clusters into two age ranges, younger than 30 Myr and older. The shaded area indicates the sector $160^\circ\,\leq\,\ell\,\leq 200^\circ$.  Panels (a), (b), and (d) allow the separation of the Perseus and Outer Arm. However, they overlap the Perseus and Local Arms. Our sample of older clusters is small and its distribution basically coincides with that of the younger clusters. These panels also suggest an extension of the Outer Arm (Cygnus arm) towards the Galactic third quadrant, which agrees with previous studies \citep{Russeil03, Pandey06, Honma07, Russeil07, Hachisuka09}. Examining panels (b) and (c) we point out that towards the anticentre most clusters are concentrated within 250 pc of the Galactic plane, which is consistent with the thin disk thickness \citep{Vallenari00, Siebert03}. However, some clusters are found at large distances from the Galactic plane, mainly in the spiral arms. The apparent limit in the distribution of clusters to $|z|\,>\,0.75$ kpc for Perseus and Local arms is a selection effect, since the FSR catalogue presents overdensities with $|b|\,\leq\,20^\circ$. On the other hand, the star cluster distribution on panels (b), (c), and (d) is consistent with the Galactic disk warp in the third quadrant and the extension of the Local arm until reaching the Outer arm \citep{Momany04, Carraro05, Moitinho06, Vazquez08}.

The gap in the spatial distribution of clusters between Perseus and Outer arms (Fig.~\ref{fig:14}) is not present when OCs from WEBDA are included in the analysis (Papers I and III). However, our clusters (Papers I, II, III, and present work) are mostly ECs. They have not had enough time to move far from their birth places and because of this trace the spiral arms better than young OCs that have had time to disperse. The WEBDA sample, in turn, appears to be rather heterogeneous both from the point of view of observational approaches and cluster properties. ECs are a minority in the WEBDA database. Optical attempts of studying ECs are often limited by dust absorption, and little capacity to separate PMS stars from field contamination, thus, in general they cannot be studied in the optical. Most open clusters in WEBDA were analysed in the optical bands and were not submitted to a field decontamination procedure. The decontamination algorithm has proved more robust to obtain fundamental parameters. Therefore, we included in Fig.~\ref{fig:14}  a homogeneous sample of clusters analysed in the 2MASS IR bands and subjected to the same analysis method.

\subsection{Age distribution}
\label{sec:7.3}

\begin{figure}
\resizebox{\hsize}{!}{\includegraphics{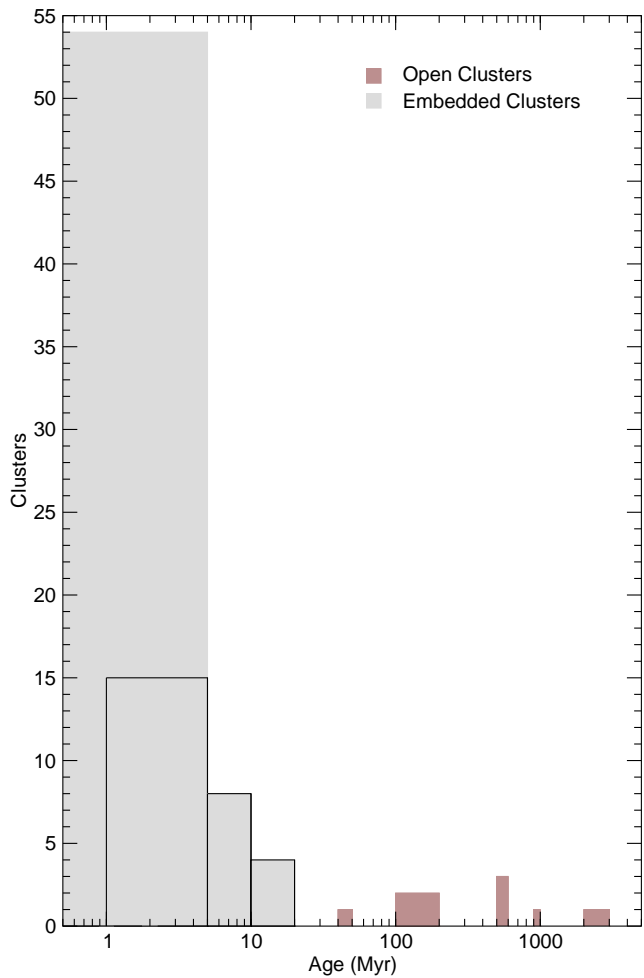}}
\caption[]{The brown-shaded histogram presents the age distribution for OCs and the grey-shaded histogram the same for ECs. The black line histogram presents ECs with evidence to be in the transition phase from embedded to open cluster.}
\label{fig:15}
\end{figure}

ECs are the embryonic phase of the cluster evolution \citep{Camargo11}. However,
as a consequence of the gas expulsion by the feedback from massive stars \citep{Tutukov78, Lada03}, most of them dissolve completely on a timescale shorter than $\approx30$ Myr
(infant mortality). After this disruptive phase, the survivors reach the maturity as OCs,
but the infant weight loss may exceed 50\% \citep{Kroupa02, Weidner07, Goddard10}. Fig.~\ref{fig:15} shows histograms with the age distribution of
the clusters in Fig.~\ref{fig:14}. We argue that some ECs appear to be in the transition phase from
ECs to OCs. These clusters are generally weakly embedded in the gas with decontaminated CMDs presenting better-defined cluster sequences and structures that follow King’s profiles. We point out that the WISE survey made it more accurate to infer the embedded nature of young star clusters (e.g. Fig.~\ref{fig:12}). Typically, at 10 Myr we
are still detecting the systematic presence of dust emission (and certainly gas), at a phase
where gas expulsion is certainly taking place. The histograms in Fig.~\ref{fig:15} show that our contribution is significant, especially for the ECs.
They also suggest that the deeply embedded phase probably does not last longer than 5 Myr,
which agrees with previous estimates \citep{Lada03, Allen07, Hetem12}. However, the complete gas expulsion occurs after 30 - 40 Myr.

\section{Concluding remarks}
\label{sec:8}

We investigate the nature of 27 star cluster candidates. Out of this total 17 are located in the anticentre region in the sector $160^\circ\,\leq\,\ell\,\leq 200^\circ$, 5 in the second, and 5 in the third Galactic quadrants, outside the anticentre sector. Twenty objects were confirmed as star clusters and we derived their parameters. We suggest deeper photometry for 4 uncertain cases and the remaining 4 are probably field fluctuations.

In addition, we provide a partial census of the open clusters towards the Galactic anticentre with fundamental parameters for 74 and structural parameters for 33 clusters. Fifteen of them are newly found clusters as a result of our infrared survey for new cluster candidates on the 2MASS database (CBB 1 to CBB 16).

The use of Embedded Clusters and their derived distances appear to  trace spiral arms with unprecedented accuracy, since they are in general embedded in the nebulae used themselves in the literature to trace spiral arms from the kinematics. Early attempts using ´´young´´ clusters or ´´young´´ open cluster distances  possibly  used relatively older clusters that may have strayed from their formation loci. 
Then, based on the Galactic distribution of these clusters, we confirm that the Outer arm extends along the second and third Galactic quadrants with Galactocentric distance in the range of $12.5-14.5$ kpc for $R_{\odot}=7.2$ kpc or $13.5-15.5$ kpc for $R_{\odot}=8.0$ kpc. Some ECs are found at large distances from the Galactic plane ($\approx2$ kpc).

We also point out that part of the unusual properties of IC 10 may be explained by a contamination of our own Galaxy in terms of stellar and nebular dust emissions, since there is a Galactic H II region related with the EC FSR 486, in the same direction. 

We suggest new steps for a detailed classification of young clusters.

\vspace{0.8cm}

\textit{Acknowledgements}: We thank an anonymous referee for significant comments and suggestions. This publication makes use of data products from the Two Micron All Sky Survey, which is a joint project of the University of Massachusetts and the Infrared Processing and Analysis Centre/California Institute of Technology, funded by the National Aeronautics and Space Administration and the National Science Foundation. We acknowledge support from INCT-A and CNPq (Brazil).

\appendix

\section{Cross-identification}
\label{sec:apendiceA}

In Table~\ref{tab2} are shown the cross-identifications or alternative designations of the clusters.

\begin{table}
\caption{Cross-identification of the clusters.}
\renewcommand{\tabcolsep}{4.2mm}
\renewcommand{\arraystretch}{1.3}
\begin{tabular}{lrrrrr}
\hline
\hline
Desig\#1&Desig\#2&Desig\#3&Desig\#4\\
($1$)&($2$)&($3$)&($4$)\\
\hline
NGC1807&Mel29&Cr61&OCL462\\
NGC1857&OCL428&&\\
NGC2331&Cr\,126&OCL-475&\\
NGC1624&Cr\,53&OCL-403&BDSB67\\
\hline
\end{tabular}
\label{tab2}
\end{table}

\section{Online material}
\label{sec:catalogue}

\begin{table*}
{\footnotesize
\begin{center}
\caption{Derived fundamental parameters for clusters in Papers I, II, and III.}
\renewcommand{\tabcolsep}{1.0mm}
\renewcommand{\arraystretch}{1.0}
\begin{tabular}{lrrrrrrrrrr}
\hline
\hline
Cluster&phase&$\alpha(2000)$&$\delta(2000)$&$\aV$&Age&$d_{\odot}$&$R_{GC}$&$x_{GC}$&$y_{GC}$&$z_{GC}$\\
&&(h\,m\,s)&$(^{\circ}\,^{\prime}\,^{\prime\prime})$&(mag)&(Myr)&(kpc)&(kpc)&(kpc)&(kpc)&(kpc)\\
($1$)&($2$)&($3$)&($4$)&($5$)&($6$)&($7$)&($8$)&($9$)&($10$)&(11)\\
\hline
BDSB\,71 &EC&05:40:51.0&35:38:20.0&$3.77\pm1.0$&$3\pm2$&$2.0\pm0.4$&$9.16\pm0.43$&$-9.02\pm0.43$&$+0.19\pm0.04$&$+0.10\pm0.03$\\
BDSB\,72 &EC&05:40:54.0&35:40:22.0&$3.77\pm1.0$&$3\pm2$&$2.1\pm0.5$&$9.26\pm0.49$&$-9.26\pm0.49$&$+0.21\pm0.05$&$+0.12\pm0.02$\\
BDSB\,73 &EC&05:40:55.0&35:44:08.0&$3.77\pm1.0$&$3\pm2$&$2.1\pm0.5$&$9.32\pm0.49$&$-9.32\pm0.50$&$+0.22\pm0.05$&$+0.12\pm0.03$\\
BPI 14&EC&05:29:00.0&34:24:00.0&$1.98\pm0.2$&$1\pm1$&$2.7\pm0.3$&$9.90\pm0.30$&$-9.89\pm0.30$&$+0.30\pm0.03$&$-0.01\pm0.01$\\
CBB\,1 &EC\textsuperscript{t}&05:39:28.0&35:40:49.0&$3.47\pm0.9$&$10\pm5$&$2.5\pm0.2$&$9.76\pm0.12$&$-9.75\pm0.12$&$+0.29\pm0.01$&$+0.11\pm0.01$\\
CBB\,2 &EC&05:40:58.0&35:42:33.0&$3.97\pm1.0$&$3\pm2$&$1.9\pm0.5$&$9.11\pm0.49$&$-9.11\pm0.49$&$+0.22\pm0.05$&$+0.09\pm0.01$\\
CBB 3 &EC&05:27:43.3&34:32:36.0&$1.98\pm0.2$&$2\pm1$&$2.7\pm0.3$&$9.90\pm0.30$&$-9.89\pm0.30$&$+0.32\pm0.03$&$-0.01\pm0.01$\\
CBB 4 &EC&05:28:29.3&34:19:50.0&$1.98\pm0.2$&$2\pm1$&$2.7\pm0.3$&$9.90\pm0.30$&$-9.89\pm0.30$&$+0.30\pm0.03$&$-0.01\pm0.01$\\
CBB 5 &EC&05:28:33.9&34:28:37.0&$1.98\pm0.2$&$2\pm1$&$2.7\pm0.3$&$9.90\pm0.30$&$-9.89\pm0.30$&$+0.31\pm0.03$&$-0.01\pm0.01$\\
CBB 6 &EC&05:29:19.0&34:14:41.4&$2.98\pm0.2$&$2\pm1$&$2.7\pm0.5$&$9.93\pm0.50$&$-9.92\pm0.50$&$+0.30\pm0.03$&$+0.0\pm0.01$\\
CBB 7 &EC&05:26:50.0&34:43:10.0&$2.98\pm0.2$&$2\pm1$&$2.5\pm0.5$&$9.69\pm0.50$&$-9.68\pm0.50$&$+0.30\pm0.03$&$-0.01\pm0.01$\\
CBB 8 &EC&05:15:50.0&34:24:00.0&$3.57\pm0.2$&$2\pm1$&$2.4\pm0.7$&$9.61\pm0.70$&$-9.60\pm0.70$&$+0.34\pm0.03$&$-0.10\pm0.01$\\
CBB 9 &EC&05:25:55.0&34:50:54.0&$3.27\pm0.2$&$2\pm1$&$2.6\pm0.5$&$9.82\pm0.50$&$-9.82\pm0.50$&$+0.33\pm0.03$&$-0.01\pm0.01$\\
Cz\,22 &OC&05:48:57.0&30:10:24.0&$1.98\pm0.1$&$200\pm50$&$2.6\pm0.1$&$9.80\pm0.20$&$-9.79\pm0.12$&$+0.03\pm0.01$&$-0.06\pm0.01$\\
FSR 734 &EC\textsuperscript{t}&05:03:22.6&42:25:15.2&$2.18\pm0.2$&$2\pm1$&$2.6\pm0.3$&$9.77\pm0.30$&$-9.74\pm0.30$&$+0.72\pm0.07$&$+0.02\pm0.01$\\
FSR\,735 &OC&04:53:52.9&40:51:42.0&$1.58\pm0.1$&$500\pm100$&$2.5\pm0.1$&$9.60\pm0.10$&$-9.56\pm0.10$&$+0.67\pm0.03$&$-0.08\pm0.01$\\
FSR 761 &EC\textsuperscript{t}&05:33:23.0&39:50:44.0&$2.78\pm0.2$&$2\pm1$&$2.5\pm0.3$&$9.73\pm0.30$&$-9.72\pm0.30$&$+0.47\pm0.04$&$+0.16\pm0.02$\\
FSR 777 &EC\textsuperscript{t}&05:27:31.0&34:44:01.0&$1.98\pm0.2$&$3\pm2$&$2.7\pm0.3$&$9.89\pm0.30$&$-9.89\pm0.30$&$+0.33\pm0.03$&$-0.01\pm0.01$\\
FSR 780 &EC&05:27:26.0&34:24:12.0&$1.98\pm0.2$&$2\pm1$&$2.7\pm0.3$&$9.90\pm0.30$&$-9.89\pm0.30$&$+0.31\pm0.03$&$-0.01\pm0.01$\\
FSR\,784 &EC\textsuperscript{t}&05:40:46.0&35:55:06.0&$3.97\pm1.0$&$5\pm2$&$2.4\pm0.5$&$9.60\pm0.10$&$-9.60\pm0.10$&$+0.27\pm0.01$&$+0.12\pm0.01$\\
FSR\,807 &EC&05:36:34.2&31:51:20.0&$5.27\pm0.1$&$5\pm3$&$1.3\pm0.1$&$8.50\pm0.10$&$-8.5\pm0.06$&$+0.08\pm0.01$&$+0.00\pm0.01$\\
FSR\,812 &EC\textsuperscript{t}&05:38:13.5&31:44:00.0&$2.48\pm0.1$&$10\pm5$&$3.3\pm0.2$&$10.5\pm0.20$&$-11.3\pm0.20$&$+0.19\pm0.01$&$+0.01\pm0.01$\\
FSR 816 &EC\textsuperscript{t}&05:39:17.0&31:30:05.0&$1.98\pm0.2$&$10\pm5$&$1.8\pm0.5$&$8.99\pm0.50$&$-8.99\pm0.50$&$+0.10\pm0.01$&$+0.01\pm0.01$\\
FSR 817 &EC&05:39:27.0&30:53:36.0&$1.98\pm0.2$&$2\pm2$&$2.3\pm0.3$&$9.56\pm0.30$&$-9.56\pm0.30$&$+0.10\pm0.01$&$0.0\pm0.01$\\
FSR\,826 &EC\textsuperscript{t}&05:52:19.1&29:55:42.7&$3.38\pm0.1$&$10\pm5$&$2.1\pm0.1$&$9.30\pm0.10$&$-9.27\pm0.01$&$+0.01\pm0.01$&$-0.02\pm0.01$\\
FSR 833 &EC&06:05:15.0&30:47:55.0&$1.79\pm0.2$&$3\pm2$&$2.9\pm0.4$&$10.1\pm0.40$&$-10.1\pm0.40$&$-0.03\pm0.01$&$+0.23\pm0.02$\\
FSR 842 &EC\textsuperscript{t}&05:34:18.8&25:36:38.0&$2.68\pm0.2$&$5\pm3$&$1.9\pm0.2$&$9.17\pm0.20$&$-9.17\pm0.20$&$-0.05\pm0.01$&$-0.13\pm0.01$\\
FSR 846 &EC&05:48:44.0&26:22:05.0&$2.98\pm0.2$&$3\pm2$&$2.5\pm0.3$&$9.70\pm0.30$&$-9.70\pm0.30$&$-0.11\pm0.01$&$-0.03\pm0.01$\\
FSR 850 &EC\textsuperscript{t}&05:45:15.0&24:45:13.0&$2.18\pm0.2$&$10\pm5$&$2.7\pm0.5$&$9.96\pm0.50$&$-9.96\pm0.50$&$-0.17\pm0.02$&$-0.11\pm0.01$\\
FSR\,852 &OC&05:53:35.0&25:10:52.0&$0.99\pm0.1$&$1000\pm200$&$2.2\pm0.1$&$9.40\pm0.10$&$-9.43\pm0.01$&$-0.16\pm0.01$&$-0.02\pm0.01$\\
FSR 864 &EC\textsuperscript{t}&05:47:49.9&21:55:32.5&$2.48\pm0.2$&$5\pm3$&$2.9\pm0.3$&$10.1\pm0.30$&$-10.1\pm0.30$&$-0.32\pm0.03$&$-0.16\pm0.02$\\
FSR 868 &EC\textsuperscript{t}&05:24:56.0&18:18:21.0&$2.98\pm0.2$&$5\pm3$&$2.7\pm0.3$&$9.90\pm0.30$&$-9.88\pm0.30$&$-0.30\pm0.03$&$-0.46\pm0.04$\\
FSR 888 &EC\textsuperscript{t}&06:22:13.0&23:24:33.0&$3.17\pm0.2$&$3\pm2$&$2.7\pm0.3$&$9.84\pm0.30$&$-9.83\pm0.30$&$-0.41\pm0.04$&$+0.21\pm0.02$\\
FSR 890 &EC&06:23:10.0&23:11:13.0&$3.37\pm0.2$&$3\pm2$&$2.6\pm0.3$&$9.77\pm0.30$&$-9.76\pm0.30$&$-0.41\pm0.04$&$+0.20\pm0.02$\\
FSR 893 &OC&06:13:45.0&21:32:54.0&$0.99\pm0.06$&$3000\pm1500$&$1.1\pm0.5$&$8.30\pm0.50$&$-8.31\pm0.50$&$-0.18\pm0.02$&$+0.04\pm0.01$\\
FSR\,904 &EC\textsuperscript{t}&06:07:09.1&19:01:08.2&$1.98\pm0.1$&$20\pm10$&$2.2\pm0.1$&$9.40\pm0.10$&$-9.42\pm0.10$&$-0.43\pm0.02$&$-0.03\pm0.01$\\
FSR\,941 &OC&06:21:47.3&15:44:22.7&$2.48\pm0.1$&$500\pm150$&$5.8\pm0.3$&$12.9\pm0.30$&$-12.8\pm0.30$&$-1.55\pm0.07$&$+0.08\pm0.01$\\
FSR 944 &EC&07:21:48.0&22:29:50.0&$3.17\pm0.2$&$3\pm2$&$2.4\pm0.3$&$9.50\pm0.30$&$-9.45\pm0.30$&$-0.63\pm0.06$&$+0.68\pm0.07$\\
FSR 946 &EC&06:10:58.0&14:09:30.0&$4.46\pm0.2$&$1\pm1$&$2.1\pm0.3$&$9.21\pm0.30$&$-9.19\pm0.30$&$-0.56\pm0.05$&$-0.08\pm0.01$\\
FSR 947 &EC\textsuperscript{t}&06:08:59.0&13:52:34.0&$2.38\pm0.2$&$2\pm1$&$2.9\pm0.3$&$10.1\pm0.30$&$-10.0\pm0.30$&$-0.80\pm0.08$&$-0.15\pm0.01$\\
FSR\,953 &OC&06:19:02.0&14:08:53.0&$1.49\pm0.1$&$500\pm150$&$2.6\pm0.2$&$9.80\pm0.20$&$-9.72\pm0.12$&$-0.75\pm0.04$&$-0.02\pm0.01$\\ 
FSR\,955 &EC\textsuperscript{t}&06:23:56.0&14:30:26.0&$1.58\pm0.1$&$10\pm5$&$3.7\pm0.2$&$10.8\pm0.20$&$-10.8\pm0.20$&$-1.09\pm0.05$&$+0.04\pm0.01$\\
$G173$\,Cl. &EC\textsuperscript{t}&05:39:28.0&35:40:13.0&$3.47\pm0.9$&$10\pm5$&$2.5\pm0.2$&$9.76\pm0.12$&$-9.75\pm0.12$&$+0.29\pm0.01$&$+0.11\pm0.01$\\
KKC\,11 &EC&05:41:24.0&35:47:34.0&$3.47\pm0.9$&$5\pm2$&$2.2\pm0.5$&$9.43\pm0.10$&$-9.43\pm0.10$&$+0.24\pm0.01$&$+0.11\pm0.01$\\
Kromb. 1 &EC\textsuperscript{t}&05:28:22.0&34:46:01.0&$1.98\pm0.2$&$3\pm2$&$2.7\pm0.3$&$9.89\pm0.30$&$-9.89\pm0.30$&$-0.32\pm0.03$&$-0.01\pm0.01$\\
NGC\,2234 &OC&06:29:20.0&16:45:27.0&$3.87\pm0.1$&$50\pm20$&$4.8\pm0.2$&$11.9\pm0.20$&$-11.9\pm0.20$&$-1.29\pm0.06$&$+0.24\pm0.01$\\
PCS\,2 &EC&05:39:13.0&35:45:53.0&$3.47\pm1.0$&$3\pm2$&$2.2\pm0.5$&$9.39\pm0.49$&$-9.38\pm0.49$&$+0.25\pm0.06$&$+0.09\pm0.02$\\
Sh2-232\,IR &EC&05:41:07.4&35:49:21.0&$4.96\pm1.0$&$5\pm2$&$1.9\pm0.5$&$9.13\pm0.50$&$-9.13\pm0.50$&$+0.22\pm0.06$&$+0.09\pm0.03$\\
Sh2-233\,SE &EC&05:39:09.6&35:45:10.0&$3.47\pm0.8$&$3\pm2$&$2.2\pm0.5$&$9.39\pm0.49$&$-9.38\pm0.49$&$+0.25\pm0.06$&$+0.07\pm0.02$\\
Sh2-235A &EC&05:40:53.0&35:42:15.0&$3.97\pm1.0$&$3\pm2$&$1.9\pm0.5$&$9.11\pm0.49$&$-9.11\pm0.49$&$+0.21\pm0.05$&$+0.10\pm0.02$\\
Sh2-235B &EC&05:40:51.0&35:41:55.0&$3.97\pm1.0$&$3\pm2$&$1.9\pm0.5$&$9.11\pm0.49$&$-9.11\pm0.49$&$+0.21\pm0.05$&$+0.09\pm0.01$\\
Sh2-235\,Cl. &EC&05:41:08.0&35:49:15.0&$3.77\pm1.0$&$5\pm2$&$2.0\pm0.6$&$9.16\pm0.09$&$-9.16\pm0.10$&$+0.22\pm0.01$&$+0.10\pm0.01$\\
Sh2-235\,E2 &EC\textsuperscript{t}&05:41:24.0&35:52:21.0&$3.97\pm1.0$&$5\pm3$&$2.1\pm0.5$&$9.29\pm0.10$&$-9.29\pm0.10$&$+0.23\pm0.01$&$+0.10\pm0.01$\\
Stock 8 &EC&05:28:07.0&34:25:28.0&$1.98\pm0.2$&$2\pm1$&$2.7\pm0.3$&$9.89\pm0.30$&$-9.89\pm0.30$&$+0.31\pm0.03$&$-0.01\pm0.01$\\
\hline
\end{tabular}
\begin{list}{Table Notes.}
\item Col. 2: evolutionary phase: EC - embedded cluster, OC - open cluster, and EC\textsuperscript{t} - transition cluster; Cols. 3 and 4: Optimised central coordinates; Col. 5: $A_V$ in the cluster's central region. Col. 6: age, from 2MASS photometry. Col. 7: distance from the Sun. Col. 8: $R_{GC}$ calculated using $R_{\odot}=7.2$ kpc as the distance of the Sun to the Galactic centre \citep{Bica06}. Cols. 9 - 11: Galactocentric components. 
\end{list}
\label{tab8}
\end{center}
}
\end{table*}

\begin{table*}
{\footnotesize
\begin{center}
\caption{Structural parameters for clusters in Papers I, II, and III.}
\renewcommand{\tabcolsep}{4.0mm}
\renewcommand{\arraystretch}{1.1}
\begin{tabular}{lrrrrrrr}
\hline
\hline
Cluster&$(1')$&$\sigma_{0K}$&$R_{core}$&$R_{RDP}$&$\sigma_{0K}$&$R_{core}$&$R_{RDP}$\\
&($pc$)&($*\,pc^{-2}$)&($pc$)&($pc$)&($*\,\arcmin^{-2}$)&($\arcmin$)&($\arcmin$)\\
($1$)&($2$)&($3$)&($4$)&($5$)&($6$)&($7$)&($8$)\\
\hline
Cz\,22 &$0.74$&$20.8\pm4.7$&$0.53\pm0.07$&$4.1\pm1.1$&$11.4\pm2.6$&$0.72\pm0.10$&$5.5\pm1.5$\\
FSR 734 &$0.76$&$24.6\pm5.4$&$1.05\pm0.21$&$7.2\pm1.4$&$14.2\pm3.1$&$1.38\pm0.28$&$9.5\pm1.5$\\
FSR\,735 &$0.71$&$19.2\pm3.2$&$0.53\pm0.06$&$6.1\pm0.7$&$9.8\pm1.6$&$0.75\pm0.09$&$8.5\pm1.0$\\
FSR 761 &$0.74$&$10.2\pm1.8$&$0.50\pm0.07$&$3.7\pm0.7$&$5.6\pm1.0$&$0.68\pm0.10$&$5.3\pm1.0$\\
FSR 777 &$0.78$&$16.2\pm2.1$&$0.59\pm0.06$&$4.3\pm0.8$&$9.8\pm1.3$&$0.76\pm0.08$&$5.5\pm1.0$\\
FSR\,784 &$0.69$&$56.7\pm4.0$&$0.25\pm0.01$&$2.1\pm0.7$&$27.0\pm1.9$&$0.36\pm0.02$&$3.0\pm1.0$\\
FSR\,807 &$0.37$&$68.2\pm10.8$&$0.15\pm0.01$&$1.5\pm0.4$&$9.4\pm1.5$&$0.41\pm0.04$&$4.0\pm0.5$\\
FSR\,812 &$0.96$&$9.9\pm3.9$&$0.70\pm0.08$&$3.8\pm1.0$&$9.1\pm3.6$&$0.72\pm0.20$&$4.0\pm1.0$\\
FSR 817 &$0.68$&$14.4\pm4.1$&$0.45\pm0.07$&$4.1\pm1.0$&$6.7\pm1.9$&$0.67\pm0.10$&$6.0\pm1.5$\\
FSR\,826 &$0.59$&$35.8\pm8.6$&$0.35\pm0.06$&$3.0\pm0.9$&$12.5\pm3.0$&$0.59\pm0.10$&$5.0\pm1.5$\\
FSR 842 &$0.62$&$12.0\pm0.5$&$0.65\pm0.06$&$4.3\pm0.1$&$4.6\pm0.2$&$1.05\pm0.09$&$7.0\pm2.0$\\
FSR 846 &$0.72$&$11.6\pm4.1$&$0.43\pm0.10$&$3.2\pm1.0$&$6.0\pm2.1$&$0.60\pm0.15$&$4.5\pm1.5$\\
FSR 850 &$0.79$&$5.5\pm1.1$&$1.20\pm0.03$&$7.9\pm2.4$&$3.4\pm0.7$&$1.50\pm0.26$&$10.0\pm3.0$\\
FSR\,852 &$0.64$&$42.5\pm18.7$&$0.60\pm0.20$&$6.4\pm0.6$&$17.4\pm7.7$&$1.01\pm0.27$&$10.0\pm1.0$\\
FSR 864 &$0.84$&$15.1\pm0.9$&$0.40\pm0.06$&$5.0\pm0.8$&$12.6\pm0.7$&$0.48\pm0.07$&$6.0\pm1.0$\\
FSR 868 &$0.79$&$8.2\pm3.4$&$0.48\pm0.15$&$3.9\pm1.6$&$5.1\pm2.1$&$0.61\pm0.19$&$5.0\pm2.0$\\
FSR 888 &$0.80$&$10.3\pm3.9$&$0.49\pm0.20$&$3.2\pm0.8$&$6.6\pm2.5$&$0.62\pm0.20$&$4.0\pm1.0$\\
FSR 890 &$0.75$&$10.1\pm3.9$&$0.36\pm0.10$&$2.2\pm0.8$&$5.7\pm2.2$&$0.49\pm0.20$&$3.0\pm1.5$\\
FSR 893 &$0.32$&$36.1\pm3.9$&$0.51\pm0.08$&$2.9\pm0.6$&$3.7\pm0.4$&$1.61\pm0.24$&$9.0\pm2.0$\\
FSR\,904&$0.65$&$6.8\pm1.0$&$1.71\pm0.27$&$11.0\pm2.0$&$2.9\pm0.4$&$2.63\pm0.41$&$17.0\pm3.0$\\
FSR\,941&$1.68$&$1.9\pm1.1$&$1.44\pm0.70$&$16.0\pm3.4$&$5.5\pm3.3$&$0.86\pm0.42$&$9.5\pm3.0$\\
FSR 944 &$0.70$&$13.9\pm2.6$&$0.39\pm0.07$&$3.8\pm1.1$&$6.8\pm1.3$&$0.56\pm0.10$&$5.5\pm1.5$\\
FSR\,953 &$0.75$&$9.7\pm5.7$&$0.77\pm0.30$&$8.3\pm1.1$&$5.5\pm3.2$&$1.03\pm0.42$&$11.0\pm1.5$\\
FSR\,955&$1.08$&$18.0\pm3.3$&$0.39\pm0.05$&$3.2\pm1.1$&$21.0\pm3.9$&$0.35\pm0.05$&$3.0\pm1.0$\\
NGC\,2234 &$1.40$&$4.0\pm1.8$&$0.78\pm0.2$&$5.6\pm1.4$&$7.9\pm3.5$&$0.56\pm0.17$&$4.0\pm1.0$\\
Sh2-235\,Cl. &$0.56$&$138.1\pm27.4$&$0.10\pm0.01$&$1.4\pm0.3$&$43.3\pm8.6$&$0.18\pm0.02$&$2.5\pm0.5$\\
Sh2-235\,E2 &$0.60$&$71.7\pm25.0$&$0.13\pm0.03$&$1.2\pm0.3$&$25.8\pm9.0$&$0.21\pm0.05$&$2.0\pm0.5$\\

\hline
\end{tabular}
\begin{list}{Table Notes.}
\item Col. 2: arcmin to parsec scale. To minimise degrees of freedom in RDP fits with the King-like profile (see text), $\sigma_{bg}$ was kept fixed (measured in the respective comparison fields) while $\sigma_{0}$ and $R_{core}$ were allowed to vary. 
\end{list}
\label{tab9}
\end{center}
}
\end{table*}

\label{lastpage}

\begin{thebibliography}{}


\bibitem[\protect\citeauthoryear{Allen et al.}{2007}]{Allen07} 
   Allen, L. E., Mergeath, S. T., Gutermuth, R., Myers, P.C., Wolk, S., Adams, F. C., Muzerolle, J., \& Pipher, J. L. 2007, Protostars and Planets V, 361


\bibitem[\protect\citeauthoryear{Alter, Ruprecht \& Vanysek}{1970}]{Alter70} 
   Alter, G., Ruprecht, J. \& Vanysek, V. 1970, in \textit{Catalogue of star clusters and associations and Suppl.}, 2nd Ed., Akad. Kiado, Budapest, edited by G. Alter, B. Balazs \& J. Ruprecht.




\bibitem[\protect\citeauthoryear{Balaguer-N{\'u}{\~n}ez et al.}{2004}]{Balaguer04} 
   Balaguer-N{\'u}{\~n}ez, L., Jordi, C., Galad{\'{\i}}-Enr{\'{\i}}quez, D., \& Masana, E. 2004, A\&A, 426, 827

\bibitem[\protect\citeauthoryear{Bica et al.}{2003a}]{Bica03a} 
   Bica, E., Dutra, C. M., \& Barbuy, B. 2003a, A\&A, 397, 177

\bibitem[\protect\citeauthoryear{Bica et al.}{2003b}]{Bica03b} 
   Bica, E., Dutra, C.M, Soares, J., Barbuy, B. 2003b, A\&A, 404, 223

\bibitem[\protect\citeauthoryear{Bica \& Bonatto}{2005}]{Bica05} 
   Bica, E., Bonatto, C. 2005, A\&A, 443, 465

\bibitem[\protect\citeauthoryear{Bica et al.}{2006}]{Bica06} 
   Bica, E., Bonatto, C., Barbuy, B. \& Ortolani, S.  2006, A\&A, 450, 105

\bibitem[\protect\citeauthoryear{Bica et al.}{2008}]{Bica08} 
   Bica, E., Bonatto, C. \& Camargo, D.  2008, MNRAS, 385, 349

\bibitem[\protect\citeauthoryear{Bica \& Bonatto}{2011}]{Bica11} 
   Bica, E., \& Bonatto, C.  2011, A\&A, 530, 32

\bibitem[\protect\citeauthoryear{Bobylev}{2007}]{Bobylev07} 
   Bobylev, V. V., Bajkova, A. T., \& Lebedeva, S. V., 2007, As. Lett., 33, 720

\bibitem[\protect\citeauthoryear{Bolatto et al.}{2000}]{Bolatto00} 
   Bolatto, A.~D., Jackson, J.~M., Wilson, C.~D. \& Moriarty-Schieven, G.,  2000, ApJ, 532, 909


\bibitem[\protect\citeauthoryear{Bonatto et al.}{2006}]{Bonatto06a}
   Bonatto, C., Kerber L. O., Bica E., Santiago B. X., 2006, A\&A, 446, 121

\bibitem[\protect\citeauthoryear{Bonatto \& Bica}{2007a}]{Bonatto07a}
   Bonatto C. \& Bica E. 2007a, A\&A, 473, 445

\bibitem[\protect\citeauthoryear{Bonatto \& Bica}{2007b}]{Bonatto07b}
   Bonatto C. \& Bica E. 2007b, MNRAS, 477, 1301

\bibitem[\protect\citeauthoryear{Bonatto \& Bica}{2008}]{StrucPar}
   Bonatto C. \& Bica E. 2008, A\&A, 477, 829



\bibitem[\protect\citeauthoryear{Bonatto \& Bica}{2009}]{Bonatto09}
   Bonatto C. \& Bica E. 2009, MNRAS, 397, 1915

\bibitem[\protect\citeauthoryear{Bonatto \& Bica}{2010}]{Bonatto10} 
   Bonatto, C. \& Bica, E. 2010, A\&A, 516, 81

\bibitem[\protect\citeauthoryear{Bonatto \& Bica}{2011a}]{Bonatto11a}
   Bonatto C. \& Bica E. 2011a, MNRAS, 414, 3769

\bibitem[\protect\citeauthoryear{Bonatto \& Bica}{2011b}]{Bonatto11}
   Bonatto C. \& Bica E. 2011b, A\&A, 530, 32

\bibitem[\protect\citeauthoryear{Bonatto, Lima \& Bica}{2012}]{Bonatto12}
   Bonatto C., Lima E.~F. \& Bica E. 2012, A\&A, 540, 137

\bibitem[\protect\citeauthoryear{Borissova et al.}{2000}]{Borissova00} 
   Borissova, J., Georgiev, L., Rosado, M., Kurtev, R., Bullejos, A., Valdez-Guti{\'e}rrez, M., 2000, A\&A, 363, 130

\bibitem[\protect\citeauthoryear{Camargo et al.}{2009}]{Camargo09} 
   Camargo, D., Bonatto, C. \& Bica, E. 2009, A\&A, 508, 211

\bibitem[\protect\citeauthoryear{Camargo et al.}{2010}]{Camargo10} 
   Camargo, D., Bonatto, C. \& Bica, E. 2010, A\&A, 521, 42

\bibitem[\protect\citeauthoryear{Camargo et al.}{2011}]{Camargo11} 
   Camargo, D., Bonatto, C. \& Bica, E. 2011, MNRAS, 416, 1522

\bibitem[\protect\citeauthoryear{Camargo et al.}{2012}]{Camargo12} 
   Camargo, D., Bonatto, C. \& Bica, E. 2012, MNRAS, 423, 1940

\bibitem[\protect\citeauthoryear{Carraro et al.}{2005}]{Carraro05} 
   Carraro, G., V\'{a}zquez, R. A., Moitinho, A., \& Baume, G. 2005, ApJ, 630, L153

\bibitem[\protect\citeauthoryear{Cardelli, Clayton \& Mathis}{1989}]{Cardelli89} 
   Cardelli, J. A., Clayton, G.C. \& Mathis, J. S. 1989, ApJ., 345, 245

\bibitem[\protect\citeauthoryear{Cohen}{1979}]{Cohen79} 
   Cohen, R. J., 1979, MNRAS, 187, 839


\bibitem[\protect\citeauthoryear{Deharveng et al.}{2008}]{Deharveng08} 
   Deharveng, L., Lefloch, B., Kurtz, S., Nadeau, D., Pomar{\`e}s, M., Caplan, J. \& Zavagno, A. 2008, A\&A, 482, 585


\bibitem[\protect\citeauthoryear{Dias et al.}{2002}]{Dias02} 
    Dias, W. S., Alessi, B. S., Moitinho, A., \& L\'{e}pine, J. R. D. 2002, A\&A, 389, 871

\bibitem[\protect\citeauthoryear{Dutra, Santiago \& Bica}{2002}]{Dutra02} 
   Dutra, C.M., Santiago, B.X. \& Bica, E. 2002, A\&A, 383, 219

\bibitem[\protect\citeauthoryear{Dutra et al.}{2003}]{Dutra03} 
   Dutra, C.M., Bica, E., Soares, J., Barbuy, B. 2003, A\&A, 400, 533

\bibitem[\protect\citeauthoryear{Feigelson et al.}{2011}]{Feigelson11} 
   Feigelson E. D., Getman, K.~V., Townsley, L.~K., Broos, P.~S., Povich, M.~S., Garmire, G.~P., King, R.~R., Montmerle, T., et al., 2011, ApJS, 194, 9



\bibitem[\protect\citeauthoryear{Fernandes, Gregorio-Hetem \& Hetem}{2012}]{Fernandes12} 
   Fernandes, B., Gregorio-Hetem, J. \& Hetem, A. 2012, A\&A, 541, 95 


\bibitem[\protect\citeauthoryear{Friel}{1995}]{Friel95} 
   Friel, E.D. 1995, ARA\&A, 33, 381

\bibitem[\protect\citeauthoryear{Froebrich, Scholz \& Raftery}{2007}]{Froebrich07} 
   Froebrich, D., Scholz, A., \& Raftery, C. L.2007, MNRAS, 374, 399


\bibitem[\protect\citeauthoryear{Glushkova et al.}{2010}]{Glushkova10} 
   Glushkova, E.~V., Koposov, S.~E., Zolotukhin, I.~Y., Beletsky, Y.~V., Vlasov, A.~D. \& Leonova, S.~I., 2010, Ast. Letters, 36, 75

\bibitem[\protect\citeauthoryear{Goddard, Bastian \& Kennicutt}{2010}]{Goddard10} 
   Goddard, Q.~E., Bastian, N. \& Kennicutt, R.~C., 2010, MNRAS, 405, 857

\bibitem[\protect\citeauthoryear{Gon{\c c}alves et al.}{2012}]{Gonçalves12} 
   Gon{\c c}alves, D.~R., Teodorescu, A.~M., Alves-Brito, A., M{\'e}ndez, R.~H. \& Magrini, L., 2012, MNRAS, 425, 2557


\bibitem[\protect\citeauthoryear{G{\"u}ne{\c s} et al.}{2012}]{Gunes12} 
   G{\"u}ne{\c s}, O., Karata{\c s}, Y. \& Bonatto, C. 2012, NewA, 17, 720


\bibitem[\protect\citeauthoryear{Gutermuth et al.}{2008}]{Gutermuth08} 
   Gutermuth, R.~A., Myers, P.~C. \& Megeath et al., 2008, ApJ, 674, 336

\bibitem[\protect\citeauthoryear{Hachisuka et al.}{2009}]{Hachisuka09} 
   Hachisuka, K., Brunthaler, A., Menten, K.~M., Reid, M.~J., Hagiwara, Y. \& Mochizuki, N., 2009, ApJ, 696, 1981

\bibitem[\protect\citeauthoryear{Honma et al.}{2007}]{Honma07} 
   Honma, M., et al. 2007, PASJ, 59, 889

\bibitem[\protect\citeauthoryear{Hunter}{2001}]{Hunter01} 
   Hunter, D.~A., 2001, ApJ, 559, 225

\bibitem[\protect\citeauthoryear{Janes \& Adler}{1982}]{Janes82} 
   Janes, K., \& Adler, D. 1982, ApJS, 49, 425

\bibitem[\protect\citeauthoryear{Jose et al.}{2011}]{Jose11} 
   Jose, J., Pandey, A.~K., \& Ogura, K., et al., 2011, MNRAS, 441, 2530

\bibitem[\protect\citeauthoryear{Kharchenko et al.}{2005a}]{Kharchenko05a} 
   Kharchenko, N.V., Piskunov, A.E. R{\"o}ser, S., Schilbach, E. \& Scholz, R.-D. 2005a, A\&A, 438, 1163

\bibitem[\protect\citeauthoryear{Kharchenko et al.}{2005b}]{Kharchenko05b} 
   Kharchenko, N.V., Piskunov, A.E. R{\"o}ser, S., Schilbach, E. \& Scholz, R.-D. 2005b, A\&A, 440, 403

\bibitem[\protect\citeauthoryear{King}{1962}]{King62} 
   King, I. 1962, AJ, 67, 471


\bibitem[\protect\citeauthoryear{Koposov et al.}{2008}]{Koposov08} 
   Koposov, S. E., Glushkova, E. V. \& Zolotukhin, I.Yu. 2008, A\&A, 486, 771

\bibitem[\protect\citeauthoryear{Kroupa}{2001}]{Kroupa01} 
   Kroupa, P. 2001, MNRAS, 322, 231



\bibitem[\protect\citeauthoryear{Kroupa}{2002}]{Kroupa02a} 
   Kroupa P. 2002, MNRAS, 330, 707

\bibitem[\protect\citeauthoryear{Kroupa \& Boily}{2002}]{Kroupa02} 
   Kroupa, P. \& Boily, C.M. 2002, MNRAS, 336, 1188


\bibitem[\protect\citeauthoryear{Lada \& Lada}{2003}]{Lada03} 
   Lada, C.J. \& Lada, E.A. 2003, ARA\&A, 41, 57


\bibitem[\protect\citeauthoryear{Lyng\aa}{1987}]{Lynga87} 
   Lyng\aa, G. 1987, Computer-based Catalogue of Open Cluster Data, 5th ed., Tech. rep., Observatoire
de Strasbourg, Centre de Donn´ees Stellaires, Strasbourg


\bibitem[\protect\citeauthoryear{Marigo et al.}{2008}]{Marigo08} 
   Marigo, P., Girardi, L., Bressan, A., Groenewegen, M.~A.~T., Silva, L. \& Granato, G.~L. 2008, A\&A, 482, 883


\bibitem[\protect\citeauthoryear{Moitinho et al.}{2006}]{Moitinho06} 
   Moitinho, A., V{\'a}zquez, R.~A., Carraro, G., Baume, G., Giorgi, E.~E. \& Lyra, W. 2006, MNRAS, 368, 77

\bibitem[\protect\citeauthoryear{Momany et al.}{2004}]{Momany04} 
   Momany Y., Zaggia S. R., Bonifacio P., Piotto G., De Angeli F., Bedin L. R., Carraro G., 2004, A\&A, 421, L29

\bibitem[\protect\citeauthoryear{Momany et al.}{2006}]{Momany06} 
   Momany, Y., Zaggia, S., Gilmore, G., et al. 2006, A\&A, 451, 515


\bibitem[\protect\citeauthoryear{Pandey, Sharma \& Ogura}{2006}]{Pandey06} 
    Pandey, A.~K., Sharma, S. \& Ogura, K., 2006, MNRAS, 373, 255


\bibitem[\protect\citeauthoryear{Parmentier \& Pfalzner}{2012}]{Parmentier12} 
   Parmentier, G. \& Pfalzner, S. 2012, Arxiv:1211.1383v1


\bibitem[\protect\citeauthoryear{Pavani et al.}{2011}]{Pavani11} 
   Pavani, D.~B., Kerber, L.~O., Bica, E. \& Maciel, W.~J. 2011, MNRAS, 412, 1611


\bibitem[\protect\citeauthoryear{Pfalzner}{2009}]{Pfalzner09} 
   Pfalzner, S. 2009, A\&A, 498, 37


\bibitem[\protect\citeauthoryear{Santos-Silva \& Gregorio-Hetem}{2012}]{Hetem12} 
    Santos-Silva, T. \& Gregorio-Hetem, J., 2012, A\&A, 547, 107


\bibitem[\protect\citeauthoryear{Piskunov et al.}{2006}]{Piskunov06} 
   Piskunov A. E., Kharchenko N. V., R\"{o}ser S., Schilbach E., Scholz R.-D., 2006, A\&A, 445, 545


\bibitem[\protect\citeauthoryear{Russeil}{2003}]{Russeil03} 
    Russeil, D., 2003, A\&A, 397, 133

\bibitem[\protect\citeauthoryear{Russeil, Adami \& Georgelin}{2007}]{Russeil07} 
    Russeil, D., Adami, C. \& Georgelin, Y.~M., 2007, A\&A, 470, 161


\bibitem[\protect\citeauthoryear{Sanna et al.}{2008}]{Sanna08} 
   Sanna et al., 2008, ApJL, 688, 69


\bibitem[\protect\citeauthoryear{Sanna et al.}{2010}]{Sanna10} 
   Sanna et al., 2010, ApJL, 722, 244


\bibitem[\protect\citeauthoryear{Skrutskie et al.}{2006}]{Skrutskie06} 
   Skrutskie, M.F. et al. 2006, AJ, 131, 1163


\bibitem[\protect\citeauthoryear{Siebert, Bienaymé \& Soubiran}{2003}]{Siebert03} 
   Siebert, A., Bienaymé, O., \& Soubiran, C. 2003, A\&A, 399, 531



\bibitem[\protect\citeauthoryear{Tadross}{2011}]{Tadross11} 
   Tadross, A. L., 2011, JKAS, 44, 1


\bibitem[\protect\citeauthoryear{Trumpler}{1930}]{Trumpler30b} 
   Trumpler, R. J., 1930, Lick Observatory Bulletin, 14, 154

\bibitem[\protect\citeauthoryear{Tutukov}{1978}]{Tutukov78} 
   Tutukov A.V. 1978, A\&A, 70, 57


\bibitem[\protect\citeauthoryear{Vallenari, Bertelli \& Schmidtobreick}{2000}]{Vallenari00} 
   Vallenari, A., Bertelli, G. \& Schmidtobreick, L. 2000, A\&A, 361, 73



\bibitem[\protect\citeauthoryear{V\'{a}squez et al.}{2008}]{Vazquez08} 
   V\'{a}zquez, R. A., May, J., Carraro, G., Bronfman, L., Moitinho, A., \& Baume, G., 2008, ApJ, 672, 930



\bibitem[\protect\citeauthoryear{Weidner et al.}{2007}]{Weidner07} 
   Weidner, C., Kroupa, P., N{\"u}rnberger, D.~E.~A., Sterzik, M.~F., 2007, MNRAS, 376, 1879


\end{thebibliography}
\end{document}